%% file: main.tex
\documentclass[journal,twoside,web]{ieeecolor}
\usepackage{tmi}

\usepackage[utf8]{inputenc}
\usepackage{siunitx}
\usepackage{comment}
\usepackage{amssymb}
\usepackage{amsmath,pict2e}
\usepackage{gensymb}
\usepackage{graphicx}
\usepackage{graphics}
\usepackage{lipsum}
\usepackage{times}
\usepackage{xcolor}
\usepackage{xspace}
\usepackage{textcomp}
\usepackage{bm}
\usepackage{diagbox}
\usepackage[mathscr]{euscript}
\usepackage{multirow}
\usepackage{epstopdf}
\usepackage{multicol}
\usepackage{cite}
\usepackage{hyperref}
\input{macro} % latex macros

\begin{document}
\title{Blind Primed Supervised (BLIPS) Learning for MR Image Reconstruction}

\author{Anish Lahiri, \IEEEmembership{Student Member, IEEE},
Guanhua Wang, \IEEEmembership{Student Member, IEEE},
\par
Saiprasad Ravishankar, \IEEEmembership{Senior Member, IEEE},
Jeffrey A Fessler, \IEEEmembership{Fellow, IEEE}
 
\thanks{This work was supported in part by
NSF Grant IIS 1838179
and
NIH Grant R01 EB023618.
}
\thanks{A. Lahiri and G. Wang are equal contributors.}
\thanks{A. Lahiri and J. Fessler are with the
Department of Electrical and Computer Engineering, University of Michigan, Ann Arbor, MI 48109, USA (emails: anishl@umich.edu, fessler@umich.edu).}
\thanks{G. Wang is with the Department of Biomedical Engineering, University of Michigan, Ann Arbor, MI 48109, USA (email: guanhuaw@umich.edu).}
\thanks{S. Ravishankar is with the
Department of Computational Mathematics, Science and Engineering and the Department of Biomedical Engineering,
Michigan State University, East Lansing, MI 48824, USA (email: ravisha3@msu.edu).}
}

%\thanks{J. Fessler is with the Department of Electrical and Computer Engineering, University of Michigan. (email: fessler@umich.edu)}
\maketitle
%\maketitle

\input{abs}

\section{Introduction}
\input{s,intro}

\section{Problem Setup and Algorithms}
\label{theory}
\input{s,algo}

\section{Experimental Framework}
\label{methods}
\input{s,methods}

%\newpage
\section{Results}
\label{results}
\input{s,result}

 % hack to solve the cyan text color problem 
\section{Discussion}
\label{dis}
\input{s,discuss}
\newpage
\section{Conclusion and Future Work}
\label{conc}
\input{s,conclusion}

\bibliographystyle{IEEEtran}
\bibliography{references}

\clearpage
\section{Supplementary Materials}
\input{sm}

\end{document}

%% file: macro.tex
\newcommand{\fref}[1] {Fig.~\ref{#1}\xspace}
\newcommand{\freff}[2] {Fig.~\ref{#1}#2\xspace}

\newcommand{\cplx}{\mathbb{C}} % jf: better R
\newcommand{\argmin}[1] {\underset{#1}{\textnormal{arg min}}}

\newcommand{\xmath}[1] {\ensuremath{#1}\xspace}
\newcommand{\blmath}[1] {\xmath{\bm{#1}}}
\newcommand{\paren}[1]{\xmath{\left( #1 \right)}}

\newcommand{\Nc} {\xmath{N_{\mathrm{c}}}}
\newcommand{\defequ}{\triangleq}

\newcommand{\D}{\xmath{\bm{D}}}
\newcommand{\db}{\xmath{\bm{d}}}
\newcommand{\e}{\xmath{\bm{e}}}
\newcommand{\z}{\xmath{\bm{z}}}
\newcommand{\Z}{\xmath{\bm{Z}}}

\newcommand{\x}{\xmath{\bm{x}}}
\newcommand{\xtrue}{\xmath{\x_{\mathrm{true}}}}
\newcommand{\y}{\xmath{\bm{y}}}

\newcommand{\Sb}{\xmath{\bm{S}}}
\newcommand{\A}{\blmath{A}}
\newcommand{\Ac}{\xmath{\A_c}}

\newcommand{\V}{\xmath{\bm{V}}}

\newcommand{\B}{\xmath{\bm{B}}}
\newcommand{\I}{\xmath{\bm{I}}}

\newcommand{\Bs}[1]{\B^{#1}}
\newcommand{\Sbs}[2]{\Sb^{#1}_{#2}}
\newcommand{\scrM}{\xmath{\bm{\mathscr{M}}}}

\newcommand{\resp}[1]{\marginpar{\textcolor{blue}{}}}
\newcommand{\red}[1] {\textcolor{black}{#1}\xspace}

\makeatletter
\newcommand{\bigcomp}{%
  \DOTSB
  \mathop{\vphantom{\sum}\mathpalette\bigcomp@\relax}%
  \slimits@
}
\newcommand{\bigcomp@}[2]{%
  \begingroup\m@th
  \sbox\z@{$#1\sum$}%
  \setlength{\unitlength}{0.9\dimexpr\ht\z@+\dp\z@}%
  \vcenter{\hbox{%
    \begin{picture}(1,1)
    \bigcomp@linethickness{#1}
    \put(0.5,0.5){\circle{1}}
    \end{picture}%
  }}%
  \endgroup
}
\newcommand{\bigcomp@linethickness}[1]{%
  \linethickness{%
      \ifx#1\displaystyle 2\fontdimen8\textfont\else
      \ifx#1\textstyle 1.65\fontdimen8\textfont\else
      \ifx#1\scriptstyle 1.65\fontdimen8\scriptfont\else
      1.65\fontdimen8\scriptscriptfont\fi\fi\fi 3
  }%
}
\makeatother

%% file: abs.tex
\begin{abstract}

This paper examines a combined supervised-unsupervised framework involving dictionary-based blind learning
and deep supervised learning for MR image reconstruction
from under-sampled k-space data.
A major focus of the work
is to investigate the possible synergy
of learned features
in traditional shallow reconstruction
using \red{adaptive} sparsity-based priors
and deep prior-based reconstruction.
Specifically, we propose a framework that
uses an unrolled network
to refine a blind dictionary learning-based reconstruction.
We compare the proposed method
with strictly supervised deep learning-based reconstruction approaches
on several datasets of varying sizes and anatomies.
We also compare the proposed method
to alternative approaches
for combining dictionary-based methods
with supervised
learning
in MR image reconstruction.
The improvements yielded by the proposed framework suggest
that the blind dictionary-based approach preserves fine image details
that the supervised approach can iteratively refine,
suggesting that the features learned using the two methods
are complementary.

\end{abstract}

\begin{keywords}
Magnetic resonance image reconstruction, deep learning, dictionary learning,
inverse problems, unrolled neural networks, sparse representations.
\end{keywords}

%% file: s,intro.tex
% s,intro

Reconstruction of images from limited measurements requires
solving an ill-posed inverse problem. In such problems, additional regularization is typically used.
Often, such regularization reflects `prior' knowledge
about the class of images being reconstructed.
Traditional regularizers exploit the sparsity of images
in some domains \cite{Lustig2007SparseImaging, Liu2015BalancedImaging},
or low-rankness
\cite{Eksioglu2016DecoupledBM3D-MRI,Jacob2020StructuredLearning}. 
Compared to using a fixed regularizer,
such as total variation (TV) or wavelet sparsity-based regularization,
data-driven or adaptive regularization has proven to be very beneficial
in several applications
\cite{Candes2010CompressedDictionaries,Wen2020ImageRegularizers,
Pratt1969HadamardCoding,Elad2007AnalysisPriors, kobler2020total, roth2005fields}. 
In this form of reconstruction,
one or more components of the regularizer,
such as a dictionary or sparsifying transform,
are learned from data adaptively,
rather than being fixed to mathematical models
like the discrete cosine transform (DCT) or wavelets.
In particular,
methods that exploit the sparsity of image patches
in a learned transform domain
or express image patches
as a sparse linear combination of learned dictionary atoms
have found widespread use in regularized MR image reconstruction
\cite{Ravishankar2011,Ravishankar2013LearningTransforms,Ravishankar2016Data-DrivenSensing,Caballero2012,Weller2016}.

A subset of this class of adaptive reconstruction algorithms
relies only upon the measurements of the image being reconstructed
to learn dictionaries or transforms, and uses no additional training data.
These methods are dubbed \textit{blind} learning-based reconstruction algorithms
or blind compressed-sensing methods \cite{lingala2013blind, gleichman2011blind}.
One advantage of patch-based dictionary-blind reconstruction algorithms is that they do not require much (or any) training data to operate,
and effectively leverage unique patterns present in the underlying data.

With the success of deep-learning-based methods
for computer vision and natural language processing,
there has also been a rise in methods that use neural networks
to ``regularize'' (often in implicit manner) MRI reconstruction problems \cite{Schlemper2019Sigma-net:Reconstruction,Ravishankar2018DeepReconstruction,%
Aggarwal2019MoDL:Problems,Schlemper2018AReconstruction, Hammernik2018LearningData}. 
Some works treat reconstruction as a domain adaptation problem
similar to style transfer and in-painting
\cite{Yang2018DAGAN:Reconstruction, Eo2018KIKI-net:Images, Lee2017DeepMRI, mardani2018deep}. 
Correspondingly, image refinement networks,
such as the U-net \cite{Ronneberger2015U-net:Segmentation},
were adopted to correct the aliasing artifacts of the under-sampled input images. 
Although such CNN-based reconstruction methods achieved improved results
compared to compressed sensing (CS) based reconstruction,
the stability and interpretability of these models
is a concern \cite{AntunOnAIb}. 

Besides improvements through algorithms,
another driving force for supervised learning-based reconstruction
is the curation of publicly available datasets for training. 
The availability of pairwise training data owing to initiatives like \cite{ZbontarFastMRI, hcp}
has further helped showcase the ability of deep learning-based algorithms
for extracting or representing image features,
and in learning richer models for image reconstruction in MR applications.
These methods, due to their reliance on pixel-wise supervision
perform exclusively \textit{supervised} learning-based reconstruction,
barring a few exceptions \cite{Quan2018CompressedLoss,Lei2019IEEETraining}.

Consequently, due to the popularity and computational efficiency of deep learning approaches
across MRI applications,
there has been a rising trend of favoring deep supervised methods
over shallower dictionary-based methods---%
perhaps because the latter methods use ``handcrafted'' priors.
%citing the `handcrafted' nature of enforcing priors on the reconstructed image
%involved in the latter, as restrictive. 
%While it may be true that 

%An underlying assumption in favoring
The rising popularity of
supervised deep learning compared to shallow blind-dictionary learning
may be based on an underlying assumption
that the features learned using relatively unrestricted
supervised deep models
%overlaps and exceeds
subsume 
those learned in a blind fashion, and other sparsity-based priors that are deemed ``handcrafted''. % using handcrafted priors.
Though supervised deep-learned regularization
may allow for the learning of richer
models in reconstructing MR images, the aforementioned assumption is largely untested.
Moreover, deep CNNs often require relatively large datasets to train well.
This paper seeks to address these issues.

This work studies the processes of blind learning-based
and supervised learning-based MRI reconstruction from under-sampled data,
and highlights the complementarity of the two approaches
by proposing a framework that combines the two 
%modes of learning
%(i.e., learning on-the-fly, and learning from big datasets) 
in a residual fashion.
We implement and compare multiple approaches
for combining %the information between
supervised and blind learning.

Our results indicate that supervised and dictionary-based blind learning may learn complementary features,
and combining both frameworks using
``BLInd Primed'' Supervised (BLIPS) learning
can significantly improve reconstruction quality.
In particular,
the combined reconstruction better preserves
fine higher-frequency details that are very important in many clinical settings.
We also find that this improvement from combining blind and supervised learning
is relatively robust to changes in training dataset size,
and across different imaging protocols.

The rest of this paper is organized as follows.
Section \ref{theory} describes the blind and supervised learning-based approaches
and the proposed strategies for combining them.
Section \ref{methods} details the experiment settings,
including datasets, hyper-parameters, and control methods.
Section \ref{results} presents the results and Section \ref{dis} provides related discussion.
Finally, Section \ref{conc} explains our conclusions and plans for future work.

%% file: s,algo.tex
% s,theory

This work combines two modern approaches to MR image reconstruction:
dictionary-based blind learning reconstruction
and CNN-based supervised learning reconstruction.
The former approach capitalizes on the sparsity of natural images in an adaptive dictionary model.
Usually, this method involves expressing patches in the MR image
as a linear combination of a small subset of atoms or columns of a dictionary. 
Across several applications, including MR image reconstruction,
%it has been observed that
learned or adaptive dictionaries often provide better representations of signals than fixed dictionaries.
When these dictionaries are learned from the image being reconstructed,
using no additional information, they are called \textit{blind},
and can be considered to be `tailored' specifically to the reconstruction at hand. 
Since individual image patches are approximated by different atoms,
overcomplete dictionaries are often preferred for this approach
because of their ability to provide richer representations of data.

\resp{1.1}\red{
For supervised learning reconstruction,
this paper uses an unrolled network algorithm
similar to the state-of-the-art method MoDL
\cite{Aggarwal2019MoDL:Problems},
whose variants have achieved top performance in recent open data-driven competitions in MR reconstruction \cite{Schlemper2019Sigma-net:Reconstruction, ramzi2020xpdnet}.
}
As `unrolled' implies, the method consists of multiple iterations or blocks.
In each iteration, a CNN-based denoiser updates the image from the previous iteration.
A subsequent data-consistency update ensures the reconstructed image is consistent
with the acquired k-space measurements.
By incorporating CNNs into iterative reconstruction,
MoDL demonstrates improved reconstruction quality and stability
compared to other direct inversion networks on large public datasets
\cite{Schlemper2019Sigma-net:Reconstruction}.

Given a set of k-space measurements
$\y_c \in \cplx^p, \ c=1, \ldots, \Nc,$
from \Nc coils with corresponding system matrices
$\Ac \in \cplx^{p \times q}, \ c=1, \ldots, \Nc$,
this section reviews the procedures of reconstruction
using blind and supervised learning,
and then proposes a method for combining them,
along with a few special cases.
We write the system matrix for the $c$th coil as
$\Ac = P \mathcal{F} \V_c$,
where $P \in \{0,1\}^{p\times q}$ incorporates
the mask that describes the sampling pattern, 
%zeroes out non-acquired % no! p < q so it selects, not zeroes!
$\mathcal{F}\in \cplx^{q\times q}$ is the Fourier transform matrix
and $\V_c \in \cplx^{q\times q}$ is the $c$th coil-sensitivity diagonal matrix,
pre-computed from fully sampled k-space
using the E-SPIRiT algorithm \cite{espirit}.

\subsection{Reconstruction using Blind Dictionary Learning}

Like most model-based regularized reconstruction approaches,
the blind sparsifying  dictionary learning-based reconstruction scheme
solves for an image \x that is consistent with acquired measurements,
and possesses properties that are ascribed to the image (or a class of images).
Mathematically, the approach optimizes 
a cost function that balances a data-fidelity term 
and a data-driven sparsity inspired regularization term as follows
\cite{Ravishankar2011MRLearning}:
\begin{equation}
    \argmin{\x} ~ \nu \sum_{c=1}^{N_c} \|\A_c \x - \y_c\|_2^2 + \mathcal{R}(\x),
    \label{eq:regrecon}
\end{equation}
where $\nu>0$ reflects confidence in data fidelity
and $\mathcal{R}(\x)$ is a regularizer that,
in the case of synthesis dictionary-based regularization,
reflects the presumed sparsity of image patches as follows:
\begin{equation}
    \begin{split}
        \mathcal{R}(\x)=\underset{\D, \Z}{\min}~&
    \sum_{j=1}^{N_1}\|\mathcal{P}_j{\x} - \D\red{\e_j}\|_2^2+\lambda^2\|\red{\e_j}\|_0
    \\&\text{s.t.}\qquad \|\db_u\|_2=1~~\forall~u,
    \end{split}
     \label{eq:dl}
\end{equation}
where $\mathcal{P}_j$ extracts the $j$th 
$\sqrt{r}\times\sqrt{r}$
overlapping patch of an image as a vector, 
$\D \in \cplx^{r\times U}$ denotes an overcomplete dicitionary, 
$\db_u$ its $u$th atom, 
\resp{4.13}\red{$\e_j$ the sparse codes for the $j$th patch and} 
the $j$th column of $\Z$,
and 
$\lambda$ is the sparsity penalty weight for dictionary learning, respectively.

A typical approach to solving this blind dictionary learning reconstruction problem
is to alternate between updating the dictionary and sparse representation
in \eqref{eq:dl}
using the current estimate of the image \x,
called \textit{dictionary learning},
and then updating the reconstructed image itself
(\textit{image update}) through \eqref{eq:regrecon}
using the current estimate of the regularizer parameters
\cite{Ravishankar2017EfficientProblems}.
This alternation between dictionary learning and image update
is repeated several times
to obtain a \textit{clean} reconstruction.
Let $\Bs{i}(\cdot)$ denote the function representing the $i$th iteration of this algorithm,
and $\x_i \in \cplx^q$ be the reconstructed image at the start of the iteration,
then we have
\begin{equation}
    \x_{i+1} = \Bs{i}(\x_i) = \B\big(\x_i;\nu_i,\lambda_i,\{\A_c,\y_c\}_{c=1}^{N_c}\big),
    \label{eq:blind}
\end{equation}

where $\nu_i,\lambda_i$ denote regularization parameters
at the $i$th iteration
for data fidelity
and for dictionary learning, respectively.
After $K$ iterations, we have,
\begin{gather}
    \x_{\text{blind}} = \x_K
    %= \bigcomp_{i=1}^{K}\Bs{i}(\x_0)
    = \paren{\bigcomp_{i=0}^{K-1}\Bs{i}}(\x_0)
    %= \Bs{K}\paren{ \ldots \Bs{2}\paren{\Bs{1}(\x_0)} }
,
\end{gather}
where $\bigcomp_{i=1}^{F}$ represents the composition of $F$ functions
$f_F\circ f_{F-1}\circ\ldots\circ f_1$,
and $\x_0$ is an inital image, possibly a zero-filled reconstruction.

In this work, we used a few iterations of the SOUP-DIL algorithm
\cite{Ravishankar2017EfficientProblems} for the dictionary and 
sparse representation update (or dictionary learning) in \eqref{eq:dl} and the conjugate gradient method
for the image update step.
\resp{4.15}\red{(See next section for details.)}

In our comparisons, we also investigated 
a similar iterative scheme as in \eqref{eq:blind}, 
but the dictionary $\D$ in \eqref{eq:dl} is not learned from data, 
and is instead fixed (e.g., to a discrete cosine transform (DCT) or wavelet basis).

\subsection{Reconstruction using Supervised Learning}

The supervised learning module (MoDL \cite{Aggarwal2019MoDL:Problems})
also aims to solve \eqref{eq:regrecon}.
Introducing an auxiliary variable \z,
\eqref{eq:regrecon} becomes:
\begin{equation}
    \argmin{\x, \z} ~ \nu \sum_{c=1}^{N_c} \|\A_c \x - \y_c\|_2^2 + \mu\|\x-\z\|_{2}^2 +  \mathcal{R}(\z),
    \label{eq:altmin}
\end{equation}
where $\mu$ controls the consistency penalty between \x and \z.
MODL updates \x and \z in alternation.
The \z update is:
\begin{align}
    \z_{l+1} &= \argmin{\z} \ \mathcal{R}(\z) + \mu \| \x_l - \z \|_2^2 
    %&
%    = (\D_\theta + \I) (\x_l) % the D here is undefined so omit
\label{z1}
.\end{align}
We replace the proximal operator in \eqref{z1}
with a residually connected denoiser $\D_\theta+\I$
applied to $\x_l$,
where \I is the identity mapping.

The \x update involves a regularized least-squares minimization problem:
\begin{equation}
    \label{x1}
    \x_l = \argmin{\x} \ \nu \sum_{c=1}^{N_c} \|\A_c \x - \y_c\|_2^2 + \mu \|\x - \z_l\|_2^2
,\end{equation}
solved via conjugate gradient method.

Similar to blind learning,
the $l$th {iteration} of supervised residual learning-based reconstruction algorithm
can be written:
\begin{equation}
\begin{split}
    \label{modleqn1}
    &\x_{l+1} = \Sbs{\theta}{l}(\x_l) =
    \Sb\big(\x_l;\nu_l, \{\A_c,\y_c\}_{c=1}^{N_c} \big)
    ,\\
    &\Sb\big(\bar{\x};\nu,\{\A_c,\y_c\}_{c=1}^{N_c}\big)
    \defequ
    \\&
    \argmin{\x}~ \nu \sum_{c=1}^{N_c} \|\A_c \x - \y_c\|_2^2
    + \|\x-\big(\D_\theta(\bar{\x})+\bar{\x}\big)\|_2^2
        ,
\end{split}
\end{equation}
\resp{2.7}\red{where $\bar{\x}$
denotes the input image for the residual learning-based reconstruction algorithm.} 
After $L$ iterations, we have
\begin{gather}
    \x_{\text{supervised}} = \x_L = \paren{\bigcomp_{l=0}^{L-1}\Sbs{l}{\theta}}(\x_0).
\end{gather}
The network parameters $\theta$ are learned in a supervised manner
so that $\x_\text{supervised}$ matches known ground truths
(e.g., in mean squared error or other metrics) on a training data set.

\subsection{Combining Blind and Supervised Reconstruction}

\begin{figure*}[h!]
    \centering
    \includegraphics[width = 0.95\textwidth]{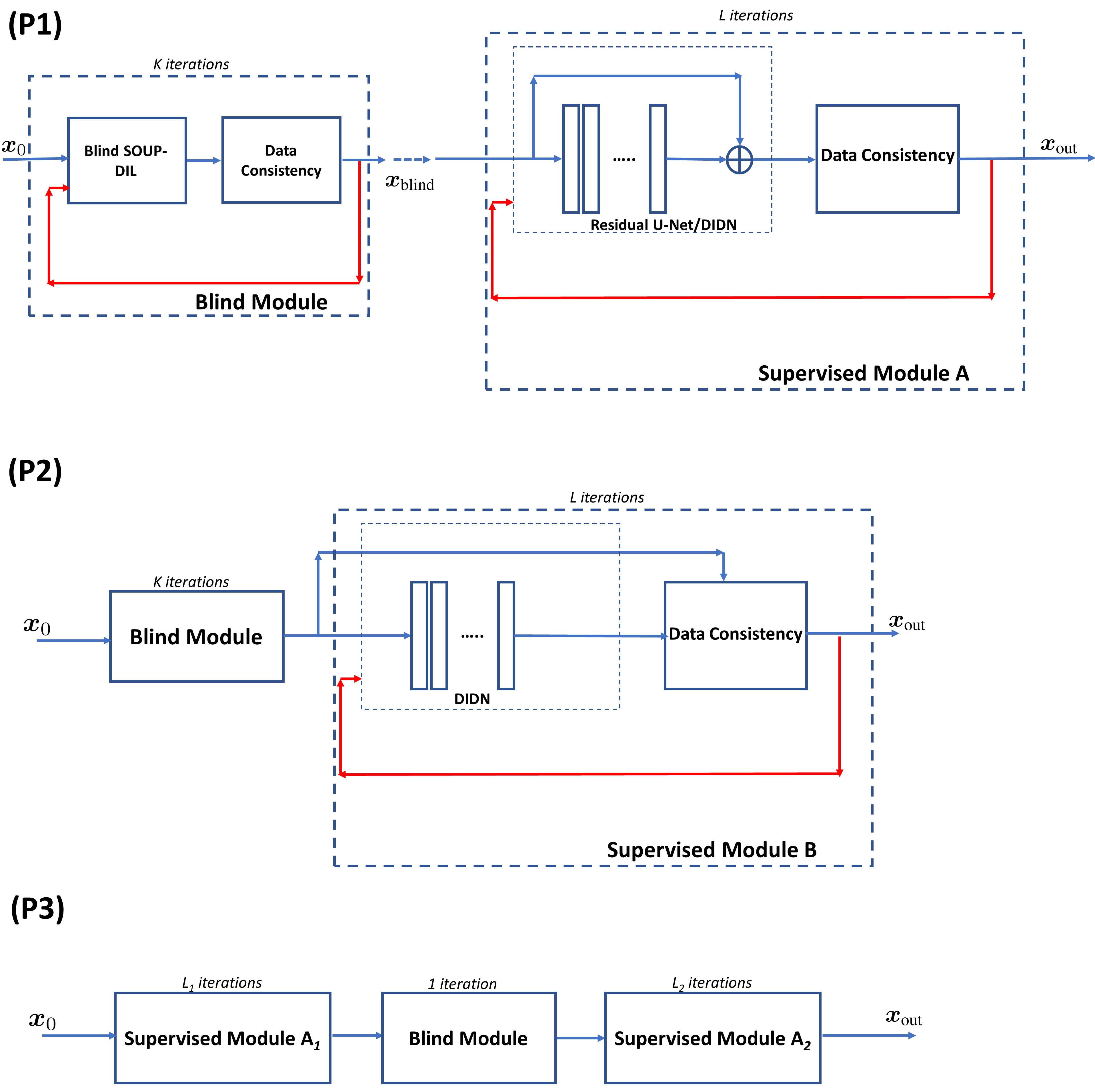}
    \caption{\red{Proposed pipelines (P1), (P2) and (P3)
    for combining blind and supervised learning-based MR image reconstruction.} }

    \label{fig:pipeline}
\end{figure*}

\fref{fig:pipeline} (P1) depicts
our proposed BLIPS approach to combining blind and supervised learning.
The skipped connection in the deep network
enables the addition of the previous iterate to the output of the denoiser
during supervised reconstruction,
and ensures {separation}
(the output of the residual denoiser gets added to the blind image going into data consistency)
of the blind learned image and the supervised learned image
in the first iteration when the aforementioned algorithms are combined.
In subsequent iterations,
{this skipped connection also} causes the denoiser to learn residual features
after the combination of blind and supervised learning in the previous iteration.
The output of the full pipeline of our proposed method \fref{fig:pipeline} (P1) is:
\begin{gather}
    \text{(P1)} \qquad
    \hat{\x} = \paren{\bigcomp_{l=0}^{L-1}\Sbs{l}{\theta}\bigcomp_{i=0}^{K-1}\B^i}(\x_0)
    \defequ \scrM_{\theta}(\x_0).
\end{gather}

\subsection{Training the Denoiser Network}

The denoiser $\D_\theta$ shares weights across iterations.
To train it, we use the output of our proposed pipeline (P1)
in a combined $\ell_1$ and $\ell_2$ norm training loss function as follows:

\begin{equation}
\begin{split}
    \hat{\theta} & =
    \argmin{\theta}~ \sum_{n=1}^{N_2}C_{\beta}(\scrM_{\theta}(\x_0^{(n)}); \x_{\text{true}}^{(n)})
    = \argmin{\theta}
    \nonumber\\ 
    &
    \sum_{n=1}^{N_2} \big( \big \|\x_{\text{true}}^{(n)} - \scrM_{\theta}(\x_0^{(n)}) \big\|_2^2
    %\\& \hspace*{8em}
    + \beta \big\| \x_{\text{true}}^{(n)} - \scrM_{\theta}(\x_0^{(n)}) \big\|_1 \big)
,
\end{split}
\label{eq:suptrn_cost}
\end{equation}
where
$n$ indexes the training data
consisting of target images $\x_{\text{true}}^{(n)}$
reconstructed from fully sampled measurements
and corresponding undersampled k-space measurements,
and
$C_{\beta}(\hat{\x}; \xtrue)$ % i didn't see this defined anywhere!
denotes the training loss function.
The initial $\x_0^{(n)}$ are obtained from the undersampled k-space measurements
using a simple analytical reconstruction such as zero-filling inverse FFT reconstruction.
Our implementation used $\beta = 0.01$ in \eqref{eq:suptrn_cost},
which was chosen empirically.

\subsection{Direct Addition of Blind and Supervised Learning}

A special case we investigate is when there is no residual connection in (P1),
and we add the blind reconstruction output directly
to the output of the supervised deep network during the data consistency update,
\resp{2.8}
\red{as described in \eqref{eqn:exp_B+S} below.
Similar to (P1), the input to the supervised module is also the blind reconstruction output.} 
We express an iteration of such an algorithm as follows:
\begin{align}
    &
    \x_{l+1} = \tilde{\Sb}_{\theta}^{l}(\x_l) =
    \tilde{\Sb}\big(\x_l;\x_{\text{blind}},\nu_l,\{\A_c,\y_c\}_{c=1}^{N_c}\big),
    \nonumber\\
    &
    \tilde{\Sb}({\x};\x',\nu,\{\A_c,\y_c\}_{c=1}^{N_c})
    = \argmin{\bar{\x}}
    \nonumber\\&
    \nu \sum_{c=1}^{N_c} \|\A_c \bar{\x} - \y_c\|_2^2+
    \|\bar{\x}-\big(\D_\theta({\x})+\x'\big)\|_2^2,
    \label{eqn:exp_B+S}
\end{align}
\resp{4.14}\red{where $\x_0=\x'$
is the initial input to the supervised module.} 
After $L$ iterations, the reconstruction is:
\begin{gather}
    \text{(P2)} \qquad
    \tilde{\x} = \paren{\bigcomp_{l=0}^{L-1}}\tilde{\Sb}^{l}_\theta(\x_{l}) = \tilde{\scrM}_{\theta}(\x_{\text{blind}}),
\end{gather}
where
$\x_{\text{blind}} = \paren{\bigcomp_{i=0}^{K-1}\Bs{i}}(\x_0)$,
\resp{4.5}
\red{as depicted in \fref{fig:pipeline} (P2).}
The training loss for this variation is:
\begin{equation}
    \hat{\theta} =
    \argmin{\theta}~ \sum_{n=1}^{N_2}C_{\beta}(\tilde{\scrM}_{\theta}(\x_\text{blind}^{(n)}); \x_{\text{true}}^{(n)}).
\end{equation}

\subsection{Combined Supervised and Blind Learning with Feedback} 

Since iterations of blind learning-based reconstruction take significantly longer
than propagating an image through a deep network,
we investigated a feedback-based pipeline that reduces computation
by only approximately optimizing the objective
of blind learning reconstruction
\resp{4.15}\red{(using an outer single iteration of the blind learning module)}
that in turn is warm-started by a supervised learning reconstruction. 
%Such blind learning introduces image-adaptive features
%into the supervised learned reconstruction,
%which could improve image quality.
The result of partial blind learning is then fed into a second stage
with a supervised deep network similar to (P1),
\resp{4.5}
\red{as depicted in \fref{fig:pipeline} (P3),}
introducing image-adaptive features
that may improve image quality.
Essentially, the output for this pipeline can be expressed as:
\begin{equation}
\label{SBS}
\begin{split}
    \red{\text{(P3)}} \qquad
    \hat{\x} &=
    \paren{\bigcomp_{l=0}^{L_2-1}\Sbs{l}{\theta_2}
    \circ \B^1 \circ
    \bigcomp_{l=0}^{L_1-1}\Sbs{l}{\theta_1}}(\x_0)
    \\ 
    & =\bar{\scrM}_{\theta_1,\theta_2}(\x_0),
\end{split}
\end{equation} 
where $\theta_1$ and $\theta_2$ are the weights
of the initial and second stage unrolled networks, respectively.
The training losses for these unrolled networks are:
\begin{equation}
\begin{split}
    \hat{\theta}_1 & = \argmin{\theta_1}~
    \sum_{n=1}^{N_2}C_{\beta}(\bar{\scrM}_{\theta_1}(\x_0^{(n)}); \x_{\text{true}}^{(n)}),
\end{split}
\end{equation}

\begin{equation}
\begin{split}
    \hat{\theta}_2 & = \argmin{\theta_2}~\sum_{n=1}^{N_2}
    C_{\beta}(\bar{\scrM}_{\theta_1,\theta_2}(\x_0^{(n)}); \x_{\text{true}}^{(n)}),
        %\hat{\theta_2}=\argmin{\theta_2}~\sum_{n=1}^{N_2} & \bigg\|\x_{\text{true}}^{(n)}-\bar{\scrM}_{\theta_1,\theta_2}(\x_{0}^{(n)})\bigg\|_2^2\\
        %&+\beta\bigg\|\x_{\text{true}}^{(n)}-\bar{\scrM}_{\theta_1,\theta_2}(\x_{0}^{(n)})\bigg\|_1,
\end{split}
\end{equation}
respectively, where
$\bar{\scrM}_{\theta_1}(\x_{0}) = \paren{\bigcomp_{l=0}^{L_1-1}\Sbs{l}{\theta_1}}(\x_0)$
and the other symbols are as explained above.
We train $\theta_{1}$ and $\theta_{2}$ separately in two stages.
The training of $\theta_2$ starts after $\theta_1$ converges.
The combination of supervised and partial blind learning could be iterated.
We worked with a two-stage network architecture
and a single iteration of blind learning optimization to keep computations low.

\iffalse
\textcolor{blue}{\item \textbf{Supervised Deep Dictionary Interpretation:}
Another special case of the above algorithm occurs
when we express the denoiser $\D_\theta$
as the composition $\D^*_{\theta_2}\circ\Z^*_{\theta_1}$, and ensure that $\D^*_{\theta_2}$,
the final few layers of $\D_\theta$ is linear (no ReLU or other activation functions),
as well as enforce sparsity on the activation of $\Z^*_{\theta_1}$ during training. 
This approach enables the interpretation of $\D^*_{\theta_2}$
as a convolutional dictionary and $\Z^*_{\theta_1}$ as a sparse coding operator. 
The training loss in this case becomes:
\begin{gather*}
    %\hat{\theta}=\argmin{\theta}~\sum_{n=1}^{N_2}\bigg\|\x_{\text{true}}^{(n)}-\paren{\bigcomp_{l=1}^{L}\Sb}\big(\D^*_{\theta_2}\circ\Z^*_{\theta_1}(\x_{K+l-1})+\x_{K+l-1};\nu_l,\{\A_c,\y_c\}_{c=1}^{N_c}\big)\bigg\|_2^2+\lambda\sum_{l=1}^{L}\bigg\|\Z^*_{\theta_1}(\x_{K+l-1})\bigg\|_1,
    \hat{\theta}=\argmin{\theta}~\sum_{n=1}^{N_2}\bigg\|\x_{\text{true}}^{(n)}-\paren{\bigcomp_{l=1}^{L}\Sbs{l}{}}\big(\D^*_{\theta_2}\circ\Z^*_{\theta_1}(\x_{K+l-1})+\x_{K+l-1}\big)\bigg\|_2^2+\lambda\sum_{l=1}^{L}\bigg\|\Z^*_{\theta_1}(\x_{K+l-1})\bigg\|_1,
\end{gather*}
where $\x_K= \paren{\bigcomp_{i=1}^{K}\Bs{i}}(\x_0)= \x_{\text{blind}}$.
\footnote{implementing on toy problem first}}
\fi

%\end{itemize}

%% file: s,methods.tex
% s,methods

\subsection{Training and Test Dataset}

We trained and tested both our method and a strict supervised learning-based method
with the same deep learning architecture (described below) on two datasets\footnote{The code will be publicly available on Github if accepted.}.
The first was a randomly selected subset from the fastMRI knee dataset,
while the second consisted of the entire fastMRI brain dataset \cite{ZbontarFastMRI}. 
In the first case, our dataset for training and testing consisted of 8705 knee images,
and were used in experiments involving the proposed pipelines in (P1) and (P2). 
We used smaller and randomly-selected subsets for our various experiments,
which is described in detail in section \ref{results}.

To test the pipeline proposed in (P3), we used the fastMRI Brain dataset,
consisting of 23220 T1 weighted images, 42250 T2 weighted images and 5787 FLAIR slices. 
For each contrast, we reserved 500 images as the test data and the rest for training and validation.

All sensitivity maps were estimated using the ESPIRiT \cite{espirit} method.
The details of the algorithms in our work are explained below.
    
\subsection{Undersampling Masks}

\begin{figure}[h!]
\begin{center}
    \begin{tabular}{ccc}
        \includegraphics[height=1.2in]{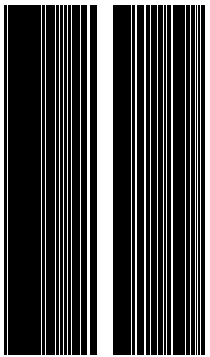}&
        \includegraphics[height=1.2in]{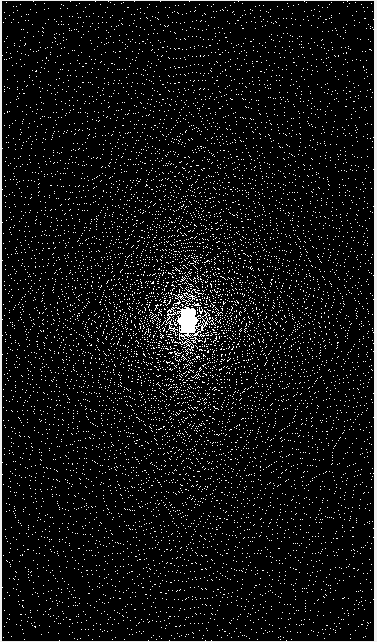}&
        \includegraphics[height=1.2in]{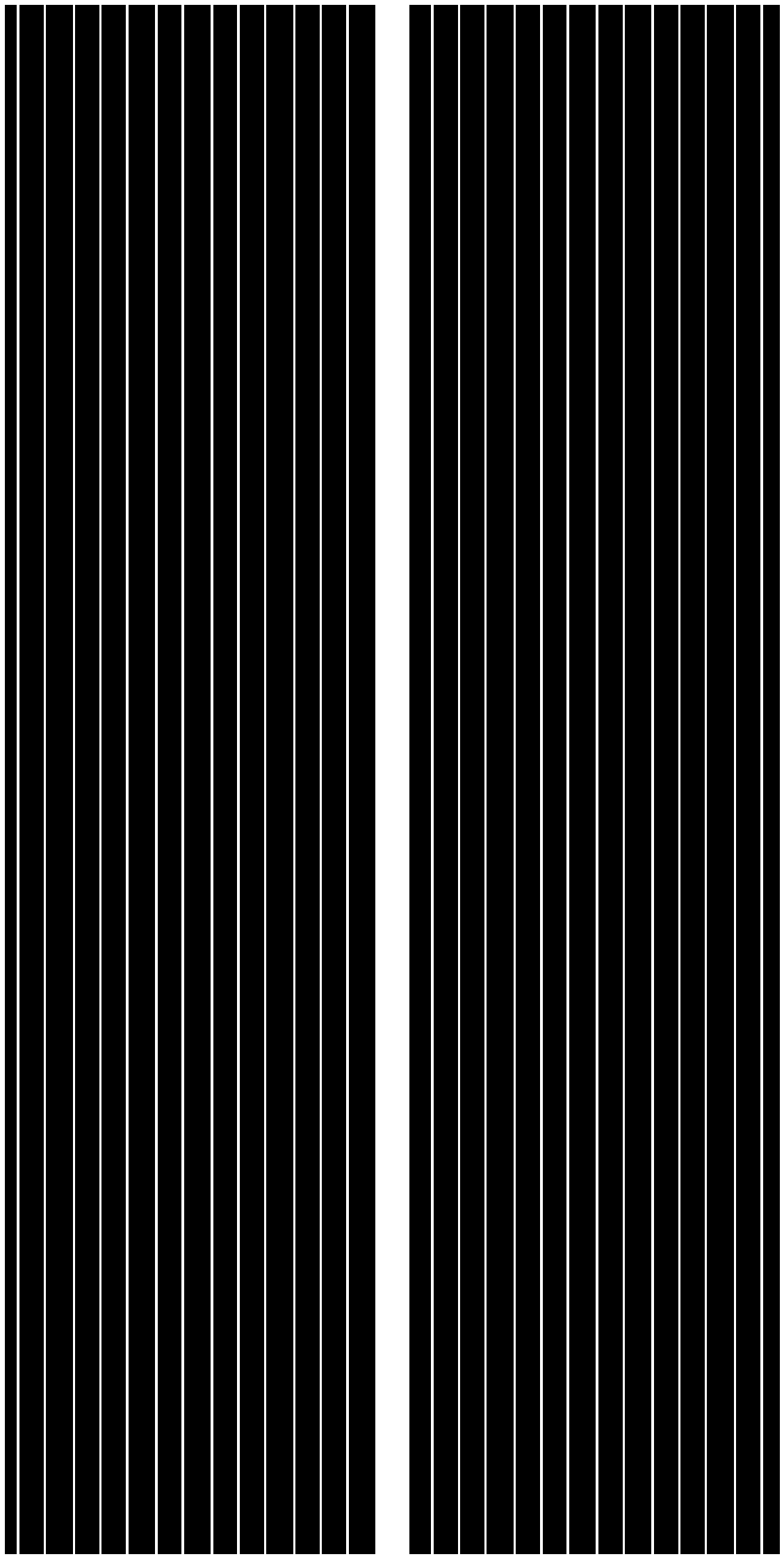}\\
        (a) & (b) & (c) \\
%        \hspace{0.1in}(a)\hspace{0.1in} & \hspace{0.1in}(b) & \hspace{0.1in}(c) \\
    \end{tabular}
\caption{
Undersampling masks used in experiments:
(a) $5$-fold undersampled 1D Cartesian phase-encoded;
(b) $20$-fold undersampled Cartesian Poisson-disk;
and (c) $8\times$ equidistant.
}
\label{fig:usml_msk}
\end{center}
\vspace{-0.1in}
\end{figure}

For experiments with the pipeline (P1),
we used three types of undersampling masks.
First, we used the $5 \times$ Cartesian phase encode undersampling mask shown in \fref{fig:usml_msk}{(a)}
that was held fixed across training and test images.
This pattern had 29 fully sampled lines in the center of the k-space, and the remaining lines were sampled uniformly at random.
We similarly tested (P1) on 2D Poisson-disk Cartesian undersampling at $20\times$ acceleration.
Finally, we tested (P1) by varying the 1D phase encode undersampling mask \freff{fig:usml_msk}{(a)} used across training and test images randomly,
\red{to further evaluate its generalizability across different sampling patterns.}
For this purpose, we used $\approx 4.5\times$ undersampling, and 24 fully sampled k-space lines.
Pipeline (P2) was tested using only the sampling pattern in \freff{fig:usml_msk}{(a)},
while (P3) was tested using $8\times$ equidistant acceleration mask shown in \freff{fig:usml_msk}{(c)}, as well as the 1D phase encode mask in \freff{fig:usml_msk}{(a)}.
This mask had 4\% fully sampled lines at the center of k-space \cite{ZbontarFastMRI}.

\subsection{Blind Dictionary Learning-based Reconstruction}
We used the SOUP-DIL algorithm \cite{Ravishankar2017EfficientProblems}
to perform blind dictionary learning-based reconstruction initialized
with a `zero-filled' reconstruction of the data. 
In both (P1) and (P2), we set the number of outer iterations to be $K = 20$,
and each outer iteration had $5$ inner iterations of dictionary learning and sparse-coding.
We set
$\nu_i = 8\times 10^{-4}$ and $\lambda_i = 0.2$ across iterations, respectively.
The {dictionary size was $36\times 144$ and the} initial dictionary was an overcomplete {inverse} DCT matrix,
while the sparse code matrix was initialized with zeros.
We used conjugate gradient method to perform the data consistent image update.
It required $\approx 170$ seconds to perform 20 iterations of SOUP-DIL reconstruction
of a single $640\times 368$ image slice, on an Intel(R) Xeon(R) E5-2698 with 40 cores.
\resp{4.15,4.1}\red{For (P3), we used only one ($K=1$) iteration of SOUP-DIL reconstruction with
$\nu = 0.5$ and $\lambda = 0.8$ on the fastMRI brain dataset
(when used on the knee dataset, these were fixed to values mentioned earlier).}
\red{For experiments involving the fastMRI brain dataset and pipeline (P1),
we only use $K=3$ outer iterations of SOUP-DIL reconstruction, due to the huge dataset size.}
A single iteration of SOUP-DIL took $\approx6.5$ seconds
to reconstruct a $640\times 320$ image on the same server.
\red{(Table VIII in the Supplementary Materials compares reconstruction time for different methods.)}
    
When performing non-adaptive dictionary-based reconstruction,
we fixed the dictionary to its inverse DCT initialization across all iterations,
while keeping all other algorithm parameters unchanged.
The experiment and results are shown in Sec. \ref{sec:fixdct} of Supplementary Materials.

\red{\resp{3.1}An additional experiment compared the compressed sensing algorithm against blind dictionary learning.
We used the MRI reconstruction instance included in the SigPy package\footnote{\url{https://github.com/mikgroup/sigpy}},
which uses the primal-dual hybrid gradient (PDHG) algorithm and 30 iterations.
The sparsity penalty is the $\ell_1$ norm of a orthogonal discrete wavelet transform,
with a weight of $10^{-7}$ compared with the data-fidelity term.}

\subsection{Supervised Reconstruction}

The denoiser $\D_\theta$ we used is the Deep Iterative Down-Up Network \cite{Yu2019DeepDenoising},
which has been shown to be efficient on previous benchmark research with the same fastMRI dataset
\cite{Schlemper2019Sigma-net:Reconstruction} 
and in an image denoising competition \cite{Abdelhamed2019NTIREResults}.
Real and imaginary component of the complex-valued images are formulated
as two input channels of the network.
The magnitude of the input image is normalized by the median absolute value.
The batch size is set to 4. We set
the data-fidelity weight $\nu = 2$ for the supervised learning.

In each iteration of \eqref{modleqn1},
we used the conjugate gradient method to solve the least-squares minimization problem. Backpropagation of the least-squares problem (calculation of the Jacobian-vector product) is also performed using the conjugate gradient method.
Here we set $L = 6$ to balance reconstruction quality and model dimension.
In the inference phase, the time cost is around 1.2s for a 20-channel $640\times 320$ slice on a single Nvidia(R) GTX1080Ti GPU. 
For a fair comparison,
the denoiser training settings are the same between different scenarios in Section \ref{results}.
The number of epochs is set to 40, with a linearly decaying learning rate from 1e-4 to 0.
The optimizer was Adam \cite{kingma2014adam},
with parameter $\beta$s = $[0.5, 0.999]$.
    
\subsection{Performance Metrics}
For a quantitative comparison of the reconstruction quality,
we used three common metrics:
peak signal-to-noise ratio (PSNR, in dB), structural similarity index (SSIM) \cite{wang2004image},
and high-frequency error norm (HFEN) \cite{Ravishankar2013LearningTransforms},
to measure the similarity between reconstructions and ground truth. {The HFEN was computed as the $\ell_2$ norm of the difference of edges between the input and reference images. Laplacian of Gaussian (LoG) filter was used as the edge detector. The kernel size was set to $15\times 15$, with a standard deviation of 1.5 pixels.}

%% file: s,result.tex
% s,result
\subsection{Comparing Blind+Supervised vs Strictly Supervised Reconstruction}

\resp{4.1}\red{Table I} compares the performance
of combined blind and supervised learning
versus strictly supervised learning on datasets of various sizes using (P1).
We used 4 training dataset sizes:
\red{1105, 2244, 4198, and 8205 slices}.
10\% for each training set was reserved for validation purposes.
The test set consisted of \red{500} different slices.
\red{Training/validation set and test set are from different subjects to avoid data leakage between slices.} \resp{4.12}

\begin{figure}
    \centering
\includegraphics[width=0.95\columnwidth]{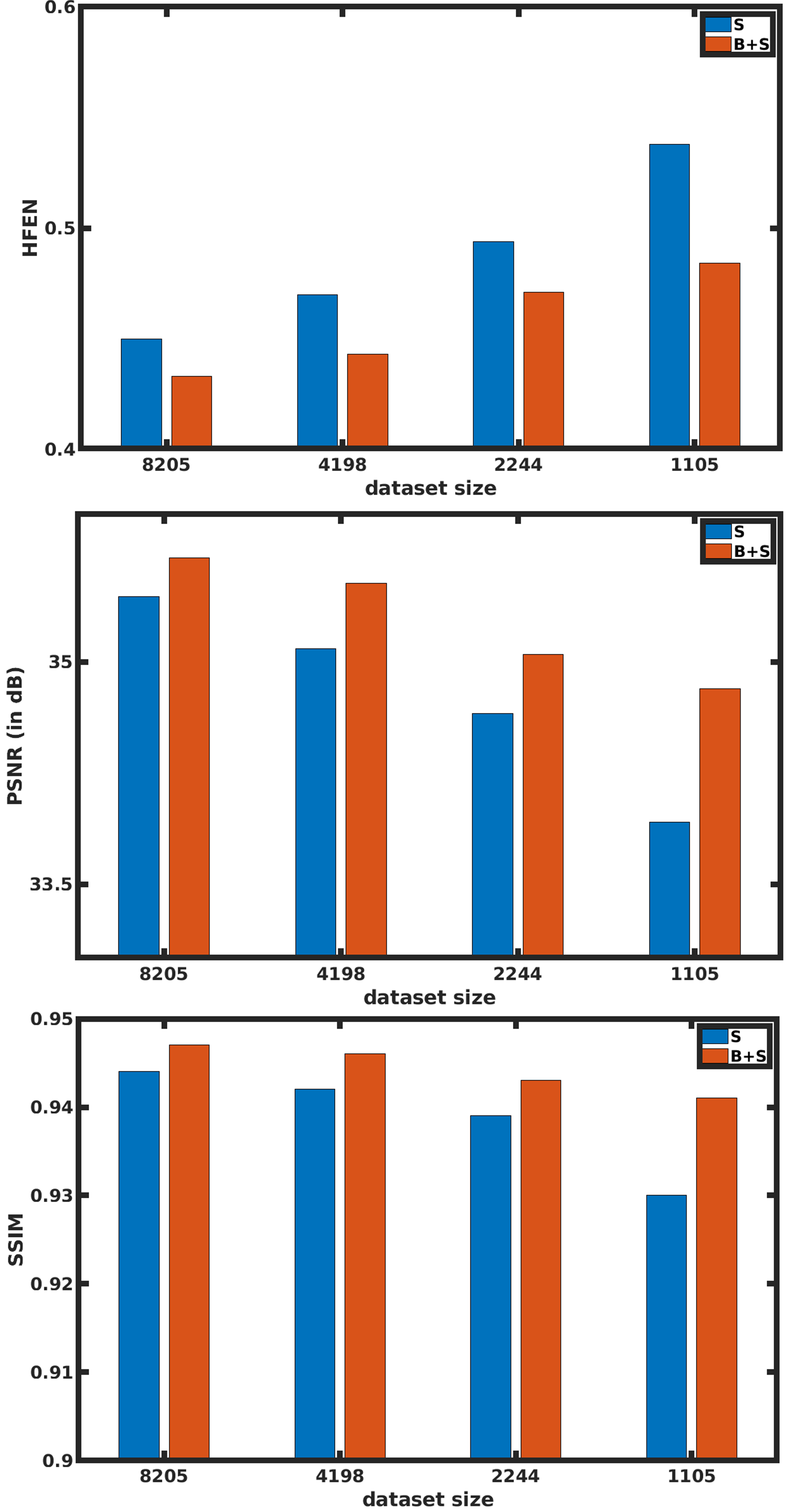}
    \caption{\red{Comparison of strict supervised learning-based reconstruction
    with BLIPS reconstruction across various knee dataset sizes.
    Table I shows the corresponding quantitative values. % is it needed? could put p-value *'s here!
    }}
    \label{fig:SvS+B_bar}
\end{figure}

Our proposed method's improvements are fairly robust
even when the total dataset size increases,
\red{\resp{2.9}
%This is further evident from
as illustrated in \fref{fig:SvS+B_bar}
that depicts Table I as a bar chart.}
Moreover, for small-scale datasets,
which are usually the case in medical imaging,
our method still provides significant improvements over the strict supervised scheme.
We conjecture that the blind learning-based reconstruction
provides an image where many artifacts have been resolved
and details have been restored
that the supervised learning reconstruction can further refine.

\red{Tables II 
and III display the quantitative results
with the 2D Poisson disk Cartesian
\resp{2.3}
sampling pattern and 1D variable density Cartesian sampling mask (changing randomly across training and test cases), respectively.
The training/validation set consisted of 4198 slices 
and the test set consisted of 500 slices (same as the 4198/500 slices in the previous case).
\resp{4.3}
The improvement provided by our scheme (B+S) over strict supervised learning (S) holds for multiple sampling masks},
and is significant under the paired t-test ($P < 0.005$).

\red{\resp{1.2}To support the assertion that BLIPS can learn
different features than supervised learning,
Figs. \ref{fig:recon:Fix_2}, \ref{fig:recon:Psn_2}, and \ref{fig:recon:Rnd_2}
also display example slices.}
Compared to supervised learning, the most obvious difference in the combined model
is the better restoration of fine details.
It can be seen that in the blind dictionary learning results,
a fair amount of fine structure is already recovered from the aliasing artifacts.
The dictionary learning results provide a foundation for supervised learning
to then residually reduce aliasing artifacts while preserving these details. 
\red{\resp{2.1,2.2}This is also strongly implied by our observations in Section VII B and accompanying \fref{fig:residue}. }

\red{Table IV
\resp{2.4,3.1}%
compares the proposed BLIPS techniques\resp{4.1} 
to strict supervised learning,
and to supervised learning initialized with compressed sensing.
The compared methods were trained and tested on identical datasets (4198 slices). 
The results indicate that the S+B+S BLIPS reconstruction yields the best performance, 
and the B+S reconstruction provides the second best performance. 
However, even compressed sensing reconstruction combined with supervised learning-based
reconstruction performs better than strict supervised learning-based reconstruction.} \\

\begin{table*}[h!]
    \centering
    \begin{tabular}{|c||c|c|c|c|c|c|c|c|}
    \hline
    Dataset Size & \multicolumn{2}{c|}{8205} & \multicolumn{2}{c|}{4198} & \multicolumn{2}{c|}{2244} & \multicolumn{2}{c|}{1105} \\ \hline
    Method  & S     & B+S            & S     & B+S            & S     & B+S            & S     & B+S            \\ \hline\hline
    SSIM     & 0.944 & \textbf{0.947} & 0.942 & \textbf{0.946} & 0.939 & \textbf{0.943} & 0.930 & \textbf{0.941} \\
    PSNR (dB) & 35.44 & \textbf{35.70} & 35.09 & \textbf{35.53} & 34.65 & \textbf{35.05} & 33.92 & \textbf{34.82} \\
    HFEN     & 0.450  & \textbf{0.433} & 0.470 & \textbf{0.443} & 0.494 & \textbf{0.471} & 0.538 & \textbf{0.484} \\ \hline
    \end{tabular}
    \begin{tabular}{|c||c|c|c|c|c|c|c|c|}
    \multicolumn{6}{l}{}\\
    \multicolumn{6}{l}{\textbf{\red{Table I}}: Comparison of %the performance of 
    supervised learning-based reconstruction (S)
    versus our proposed combined blind and supervised learning-based} \\
    \multicolumn{6}{l}{reconstruction (B+S)
    using (P1) at various knee training dataset sizes for 5$\times$ acceleration
    using 1D Cartesian undersampling. The undersampling} \\
    \multicolumn{6}{l}{mask in \freff{fig:usml_msk}{a}
    was held fixed for training and testing.
    \red{Bold digits indicate that B+S method performed significantly better than the S method }}\\
    \multicolumn{6}{l}{\red{under pairwise t-test ($P<0.005$).}}
    \end{tabular}

    \label{tab:SvBplusS_2}
\end{table*}

\begin{table}[h!]
    \centering
    \begin{tabular}{|c||c|c|c|}
    \hline
        Recon. Method & Supervised & \red{Blind} & Blind+Supervised \\\hline\hline
         SSIM & 0.960 & 0.949 & \textbf{0.963} \\
         PSNR (dB) & 35.33 & 33.38 & \textbf{35.81}\\
         HFEN & 0.434 & 0.507 &\textbf{0.403} \\
         \hline
    \end{tabular}
    \begin{tabular}{ccc}
    \multicolumn{3}{l}{} \\
    \multicolumn{3}{l}{\textbf{\red{Table II}}: Comparison of %performance of
    supervised learning-based reconstruction
    } \\
    \multicolumn{3}{l}{versus our proposed BLIPS \red{and blind learning-based} reconstruction} \\
    \multicolumn{3}{l}{ using (P1) for 20 $\times$ acceleration using Cartesian 2D Poisson disk } \\
    \multicolumn{3}{l}{undersampling
    with mask shown in \freff{fig:usml_msk}{b}. The fastMRI knee}\\
    \multicolumn{3}{l}{dataset was used for training and testing. \red{Bold digits indicate that}}\\
    \multicolumn{3}{l}{\red{B+S method performed significantly better than the S method under}}\\
    \multicolumn{3}{l}{\red{paired t-test ($P<0.005$).}}
    \end{tabular}
    \label{tab:possion}
\end{table}

\begin{table}[h!]
    \centering
    
    \begin{tabular}{|c||c|c|c|}
    \hline
        Recon. Method & Supervised & \red{Blind} & Blind+Supervised \\\hline\hline
         SSIM & 0.954 & 0.945 & \textbf{0.957} \\
         PSNR (dB) & 34.34 & 32.79 & \textbf{34.80}\\
         HFEN & 0.308 & 0.360 &\textbf{0.284} \\
         \hline
    \end{tabular}
        \begin{tabular}{ccc}
        \multicolumn{3}{l}{} \\
    \multicolumn{3}{l}{\textbf{\red{Table III}}: Comparison of performance of supervised learning-based } \\
    \multicolumn{3}{l}{reconstruction against our proposed BLIPS and \red{blind learning-based} } \\
    \multicolumn{3}{l}{reconstruction using (P1)  for  $\approx 4.5\times$ acceleration using random } \\
    \multicolumn{3}{l}{variable density 1D sampling mask (changing randomly across }\\
    \multicolumn{3}{l}{training and test cases). The fastMRI knee dataset was used for }\\
    \multicolumn{3}{l}{ training and testing. \red{Bold digits indicate that B+S method performed }}\\
        \multicolumn{3}{l}{\red{significantly better than the S method under paired t-test ($P<0.005$).}}\\
    %\multicolumn{3}{l}{\red{}}
    \end{tabular}

    \label{tab:rnd}
\end{table}

\begin{table}[h!]
    \centering
    \begin{tabular}{|c||c|c|c|c|c|}
    \hline
        Recon. Method & S & \red{B} & \red{CS+S} & B+S & \red{S+B+S}
        \\\hline\hline
         SSIM & 0.942 & 0.906 & 0.943 & {0.946} & \textbf{0.948}\\
         PSNR (dB) & 35.09 & 30.29 & 35.24 & {35.53} & \textbf{35.83}\\
         HFEN & 0.470 & 0.648 & 0.464 & {0.443} & \textbf{0.426} \\
         \hline
    \end{tabular}
    \begin{tabular}{ccc}
    \multicolumn{3}{l}{} \\
    \multicolumn{3}{l}{\textbf{\red{Table IV}}: Comparison of %performance of
    supervised learning-based reconstruction
    } \\
    \multicolumn{3}{l}{versus various proposed BLIPS reconstruction approaches using} \\
    \multicolumn{3}{l}{(P1) and (P2), and CS-initialized supervised reconstruction for}\\ 
    \multicolumn{3}{l}{5$\times$ acceleration using 1D Cartesian undersampling
    with mask} \\
    \multicolumn{3}{l}{shown in \freff{fig:usml_msk}{a}. Training was performed using 4198 knee slices}\\
    \multicolumn{3}{l}{from the fastMRI Knee dataset. \red{Bold digits indicate that S+B+S}}\\
        \multicolumn{3}{l}{\red{method performed significantly better than the S method and CS+S}}\\
    \multicolumn{3}{l}{\red{method under paired t-test ($P<0.005$).}}
    \end{tabular}
    \label{tab:compcomp}
\end{table}

\subsection{Strict Separation of Blind and Supervised Learning Reconstruction}

Table V compares explicitly combining blind and supervised learning using (P2)
without residual learning
against the proposed method for combining blind and supervised learning.
The sampling pattern here is the same as in \freff{fig:usml_msk}{a}. The dataset is the same as the 8205/500 case in Table I.
Compared to explicit consistency with blind learning results,
our latent approach reaches a better result.
The results demonstrate that rather than a fidelity prior,
the blind learned reconstruction works better as an input
to the deep residual network for further refinement.\\

\begin{table}[h!]
    \centering
    \begin{tabular}{|c||c|c|}
    \hline
        Recon. Method & Explicit & Proposed 
        \\
        & Blind + Supervised & Blind + Supervised 
        \\\hline\hline
        SSIM & 0.938 & \textbf{0.946} \\
        PSNR (dB) & 34.29 & \textbf{35.53}\\
        HFEN & 0.495 &\textbf{0.443} \\
        \hline
    \end{tabular}
        
        \begin{tabular}{ccc}
        \multicolumn{3}{l}{} \\
        \multicolumn{3}{l}{\textbf{Table V}: Comparison of 
combined blind and supervised learning using } \\
        \multicolumn{3}{l}{(P1)
versus explicit addition of blind and supervised learning using (P2)} \\
        \multicolumn{3}{l}{for the mask in \freff{fig:usml_msk}{a}. Training was performed  using 4198 knee slices}\\
       \multicolumn{3}{l}{ from the fastMRI Knee dataset. \red{Bold digits indicate that B+S method }} \\
            \multicolumn{3}{l}{\red{performed significantly better than the explicit blind + supervised method}}\\
    \multicolumn{3}{l}{\red{under paired t-test ($P<0.005$).}}
        \end{tabular}
\label{tab:SvBplusS_explicit}
\end{table}

\subsection{Combined Supervised and Blind Learning with Feedback}

For the large-scale brain dataset,
we tested the idea of using a supervised learning network's output
as a potentially improved %benign
initialization for blind learning \eqref{SBS}.
The blind learning cost is then optimized for a single iteration with this improved initialization
to incorporate additional details captured with blind learning
to improve the first supervised network's reconstruction.
The blind learning result is passed on to another (second) stage of supervised learning.
The networks' parameters $\theta_1$ and $\theta_2$ are pre-trained on all three contrasts
and fine-tuned on individual contrast, including T1w, T2w and FLAIR.
As a control method,
we concatenated two supervised learned networks sequentially,
which can also improve the reconstruction performance compared with a single unrolled supervised network, 
and demonstrates substantial improvements in PSNR, SSIM, and HFEN for S+B+S with a large dataset. 
\resp{4.1}\red{We also compare to deep supervised reconstruction
preceded by a few iterations of blind dictionary learning.
(We used 3 iterations here due to the time constraints
associated with generating data for the large fastMRI brain dataset.)}
 
\red{Table VI}\resp{2.3} summarizes the results of this comparison,
showing \red{that while S+B+S performs the best, 
even B+S}
(which on the brain dataset,
only used 3 iterations of SOUP-DIL reconstruction in the blind module)
manages to out perform strict supervised 
learning in most contrasts.
% should there be a \red here?
{\fref{fig:recon:equi_8x_brain} shows an example slice for this comparison.
Again, combined blind and supervised learning using (P3)
preserves finer details better than cascaded strict supervised learning.} \\

\begin{table*}[h!]
    \centering
    \begin{tabular}{|c||c|c|c|c|c|c|c|c|c|}
    \hline
    Dataset & \multicolumn{3}{c|}{T1w} & \multicolumn{3}{c|}{T2w} & \multicolumn{3}{c|}{FLAIR}  \\ \hline
    Method  & S+S     & S+B+S      & \red{B+S}      & S+S     & S+B+S      & \red{B+S}      & S+S     & S+B+S      & \red{B+S}                  \\ \hline\hline
    SSIM     & 0.965 & \textbf{0.968} & \red{0.966} & 0.964 & \textbf{0.967} & 0.966& 0.944 & \textbf{0.947} & 0.945 \\
    PSNR (dB) & 36.86 & \textbf{37.27} & \red{36.85} & 35.37 & \textbf{35.88} & 35.72 & 34.23 & \textbf{34.62} & 34.36 \\
    HFEN     & 0.388 & \textbf{0.369}  & \red{0.384} & 0.371 & \textbf{0.349} & 0.353 & 0.481 & \textbf{0.458}  & 0.470 \\ \hline
    \end{tabular}
    \begin{tabular}{ccc}
    \multicolumn{3}{l}{} \\
    \multicolumn{3}{l}{\textbf{\red{Table VI}}: Comparison of %the performance of
        strictly supervised learning-based reconstruction (S+S) versus the proposed combined blind and supervised learning-based} \\
    \multicolumn{3}{l}{reconstruction (S+B+S) in (P3)
        for the fastMRI brain dataset with $8\times$ undersampling with the mask in \freff{fig:usml_msk}{c}.} \\
    \end{tabular}
        \label{tab:SBSvSS}
\end{table*}

\subsection{Performance in the Presence of Planted Features}
\red{
\resp{3.2}To compare the ability of BLIPS reconstruction
and strictly supervised reconstruction
to faithfully reproduce image features
that are not present in the training dataset
(as is often the case with identifying pathologies, etc.),
we planted some} \red{features in a knee image from the fastMRI dataset,
from which raw k-space was simulated and undersampled,
inspired by
\cite{AntunOnAIb}.
The undersampling pattern was 1D variable density $\approx 4.5\times$,
and was chosen at random
to further test robustness.}

\red{\fref{fig:recon:patho}
\resp{3.2}
shows the aforementioned comparison.
The BLIPS reconstruction reproduces the planted features 
with significantly higher fidelity than strict supervised reconstruction,
and has much fewer aliasing artifacts,
as is evident from the residue maps (also pointed out by the blue arrows in the figure).}
\red{The details or edges of the planted features are better preserved} \red{in the BLIPS reconstruction
compared to strict supervised learning-based reconstruction.}
\red{The phenomena are consistent across simulated attempts we have tried.
}

%% file: s,discuss.tex
% s,discuss
This work investigated the combination of blind and supervised learning algorithms
for MR image reconstruction.
Specifically, we proposed a method that combines dictionary learning-based blind reconstruction
with model-based supervised deep reconstruction in a residual fashion.
Comparisons against strictly supervised learning-based reconstruction
indicate that the proposed reconstruction method
significantly improves reconstruction quality in terms of metrics including PSNR, SSIM, and HFEN,
across a range of undersampling and acceleration factors.
The robustness of these improvements to the training dataset size
suggests that the features learned during blind learning-based reconstruction
using a sparse dictionary adapted separately for each training and testing image
may differ significantly from features learned
by deep networks trained on a large dataset with strictly pixel-wise supervision.
{While the latter showcases the potential for removing global aliasing artifacts,
the former successfully leverages patterns in an  image that are learned just from its measurements,}
{thereby preserving the finer details of the image in the reconstruction.}
\resp{1.2}\red{This claim is further supported
by the error maps 
of regions of interest of 
reconstructed image slices.
Moreover, the experiments using planted features
suggest that BLIPS reconstruction
can adapt to, and reproduce unfamiliar} 
\red{(absent from the training set) features
better than strict supervised learning-based reconstruction. 
This ability may be a distinct benefit
in the context of identifying pathology in MRI images.}
\resp{2.1,3.1}\red{The combination of compressed sensing MRI 
and deep-supervised learning-based reconstruction
also outperformed strict supervised learning-based reconstruction, 
reinforcing that features learned using supervision
may not subsume traditional sparsity-based priors.}

Past studies have shown that deep learning-based reconstruction
is good at reducing aliasing artifacts
compared with model-based iterative methods such as compressed sensing.
The majority of supervised models are trained with pixel-wise $\ell_1/\ell_2$ norm loss.
These approaches generally produce smooth images with high PSNR but can also introduce blurring.
Other methods use GANs or perceptual loss to preserve details.
However, these data-driven methods are often known to introduce realistic artifacts,
which is very risky for medical imaging reconstruction.
In our approach, the intrinsic sparsity of MR images is exploited
%explicitly
in the dictionary learning phase
to preserve %lots of
fine structures.
Thus, our method combines the advantages of both worlds:
the representation ability of CNNs to resolve aliasing artifacts
and dictionary-based signal modeling to recover high-frequency details.
The superior performance in fine-detail recovery is reflected in the smaller HFEN values
that quantify high-frequency features. 

From the network training perspective,
compared to the pure supervised model
our network demonstrates improved stability and generalizability
since it is powered and complemented by both model-based and adaptive dictionary learning-based components.
First, on a relatively small dataset (1105/2244 images),
the method still achieved similar results as with the full (8205 images) training dataset.
This means that our method has clearly lower requirements on the amount of training data to work well compared to the massive amount of training data
needed by typical deep learning-based reconstruction algorithms.
Second, the improvements hold across different sampling patterns with very different PSFs.
\resp{3.4}\red{Third, although 40 training epochs were used in experiments,
our approach requires only 5-8 epochs to converge
(with no obvious over-fitting seen thereafter).
In contrast, the supervised model required 20-30 epochs for the training loss to converge.} 

%Comparisons with a
%non-adaptive dictionary-based initialization
%for supervised reconstruction
%reveal that the sparse synthesis dictionary model for image patches
%contributes to the improvements yielded by our proposed method.
%We find that
%The overcomplete DCT dictionary sufficed to capture the features in Knee MRI images.
%More careful tuning of hyperparameters may be required
%to fully capitalize on the benefits of dictionary learning.
%One way to achieve this would be to vary the sparsity penalty weight, $\lambda$,
%across outer-iterations of dictionary learning-based reconstruction
%as is done in \cite{Ravishankar2017EfficientProblems}.

Due to the serial nature of the SOUP-DIL algorithm \cite{Ravishankar2017EfficientProblems}
used for dictionary learning here,
our algorithm's reconstruction time is higher than that of strictly supervised reconstruction.
The computational bottleneck is in the atom-wise block-coordinate descent approach
to dictionary updating, which cannot be accelerated by simple vectorization.
These alternating updates between each dictionary atom and the corresponding sparse codes
\cite{Ravishankar2017EfficientProblems}
allow for the blind algorithm to residually learn and represent features in the reconstructed image.
Further acceleration of the blind dictionary learning approach
might be needed to use the approach in clinical settings that need
%this may restrict the application in
a real-time imaging reconstruction workflow.
However, it may be still acceptable for most conventional settings
since the scanning itself is often the throughput bottleneck. 
The proposed S+B+S approach involves a much quicker 
(partial) dictionary learning-based step compared 
to the proposed vanilla B+S approach. 
Other fast blind learning approaches involving transform learning
\cite{Ravishankar2013LearningTransforms}
could also make our schemes much more efficient.

\newcommand{\hh}{1.6in} % use macros not "2.2in" repeatedly!

\begin{figure}[h!]
\begin{center}
\begin{tabular}{cc}
\textbf{Fully Sampled} & \\ 
\includegraphics[height=\hh]{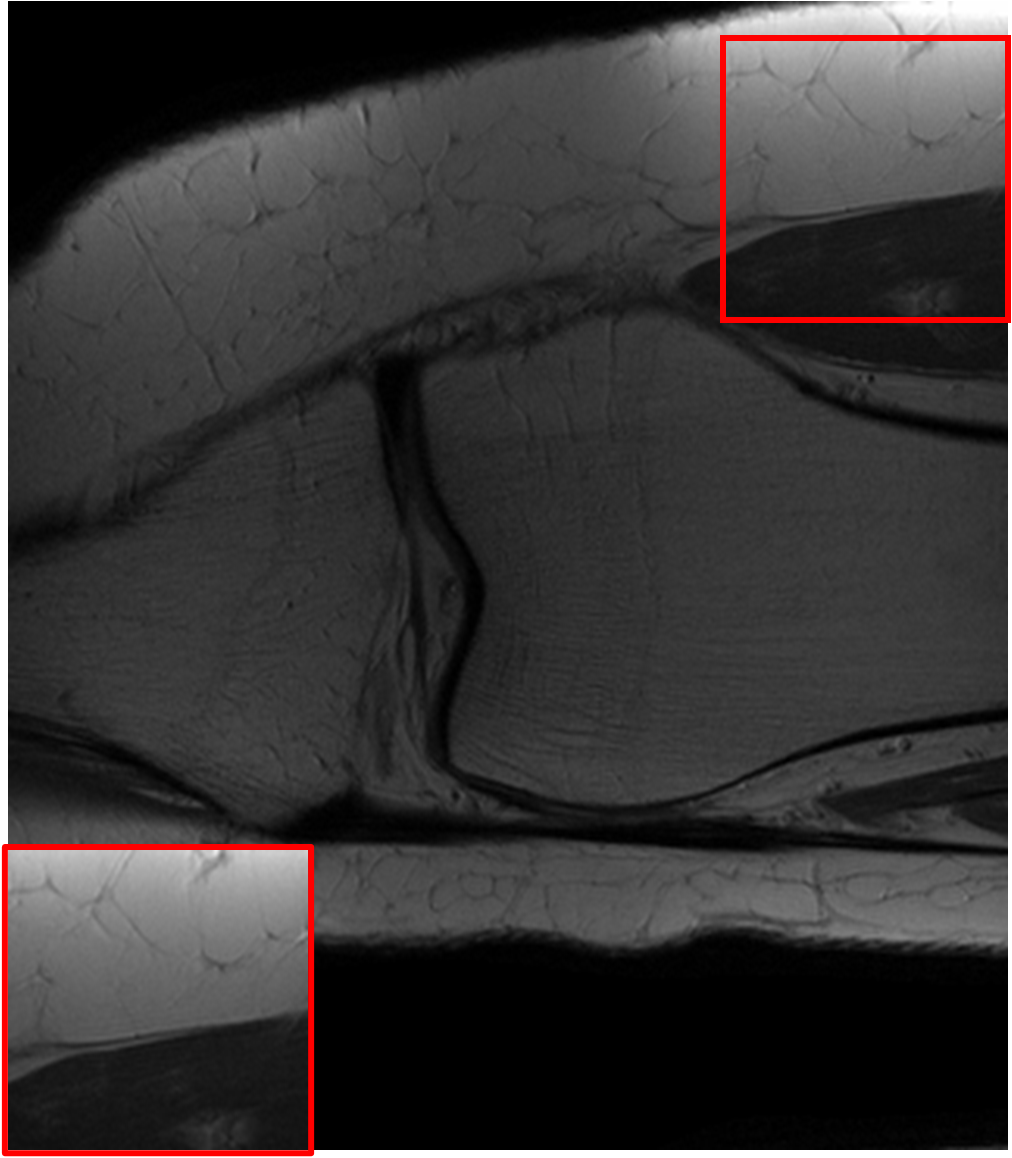} & \\
\hspace{0.1in}(a)(PSNR/SSIM/HFEN)\hspace{0.1in} & \\
\textbf{Blind+Supervised Recon.} & \textbf{Supervised Recon.} \\ 
 \includegraphics[height=\hh]{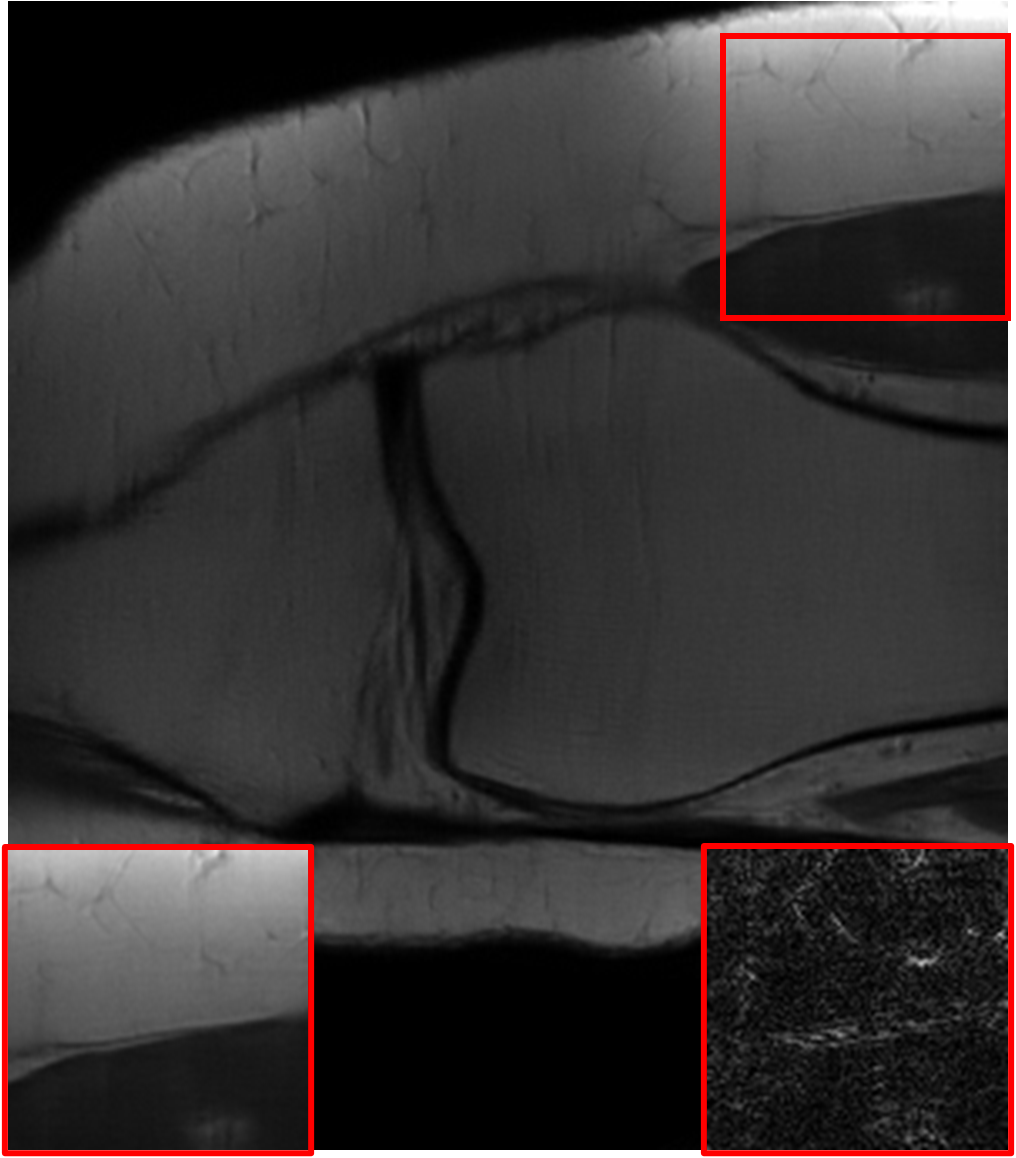}  & \includegraphics[height=\hh]{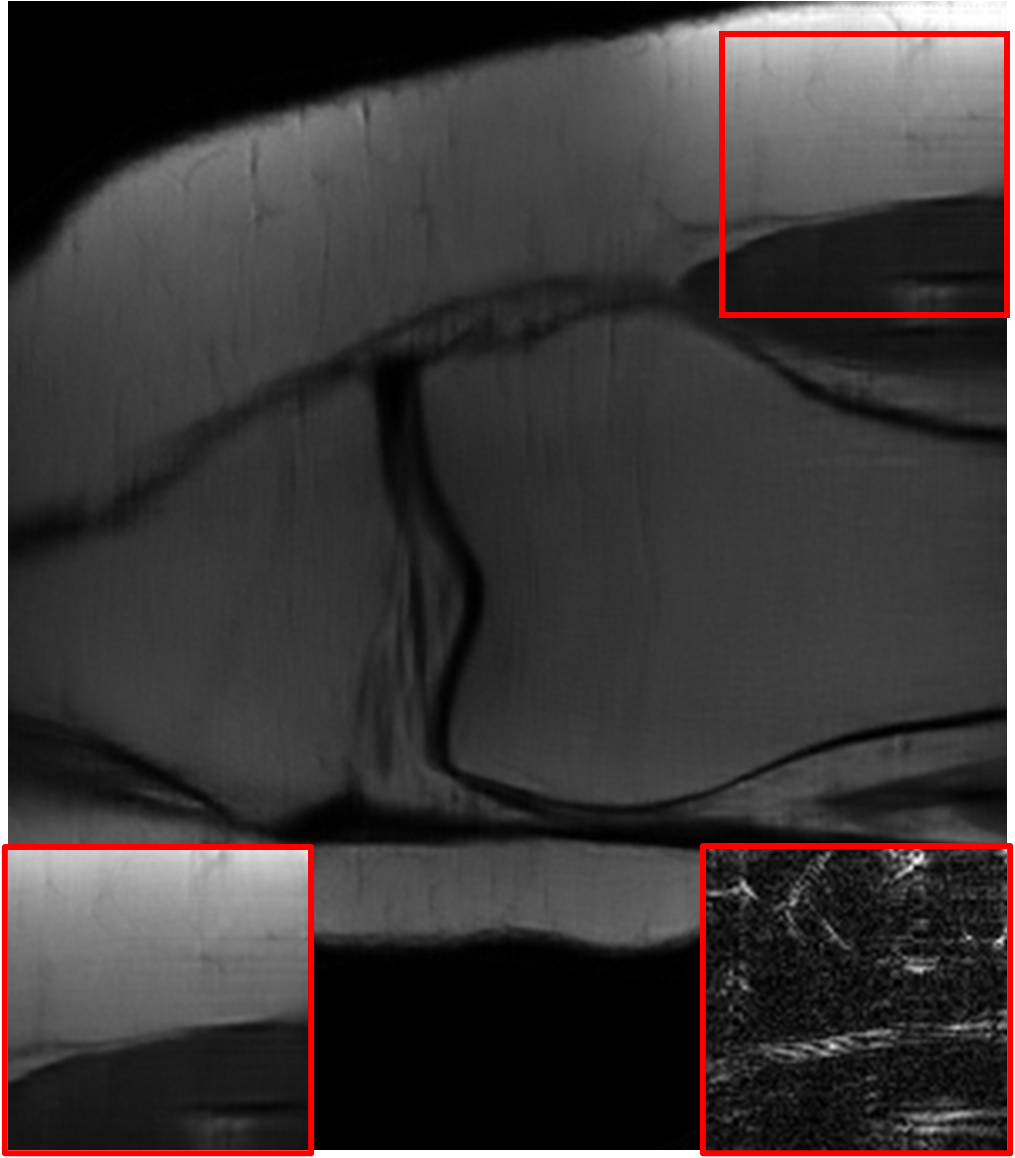}\\
 \hspace{0.1in}(b) (41.24/0.984/0.355) \hspace{0.1in} & \hspace{0.1in}(c) (39.59/0.980/0.428)\hspace{0.1in }\\
\textbf{Blind Dict. Learning Recon.} & \textbf{Zero-Filled Recon.}\\
 \includegraphics[height=\hh]{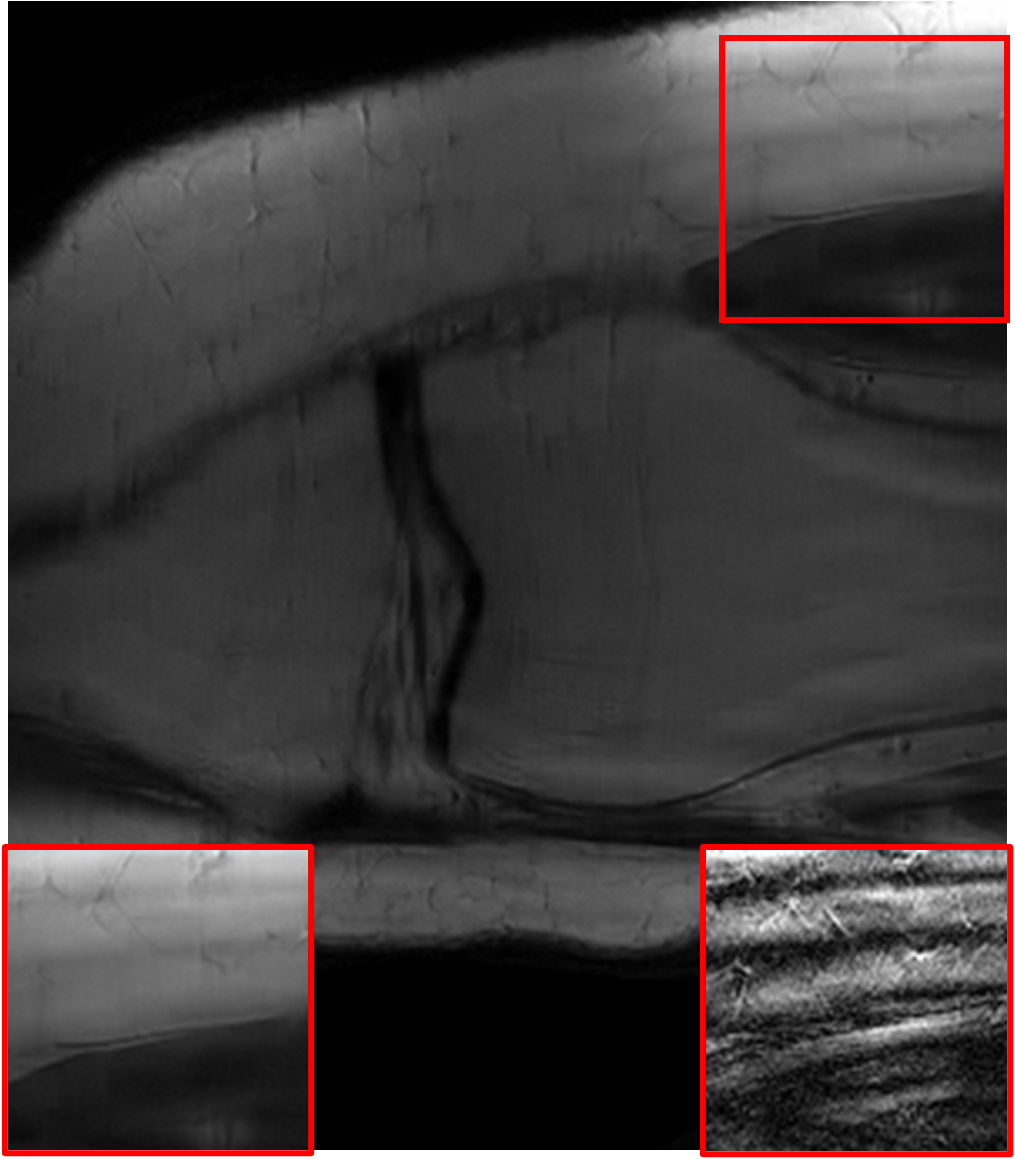}  & \includegraphics[height=\hh]{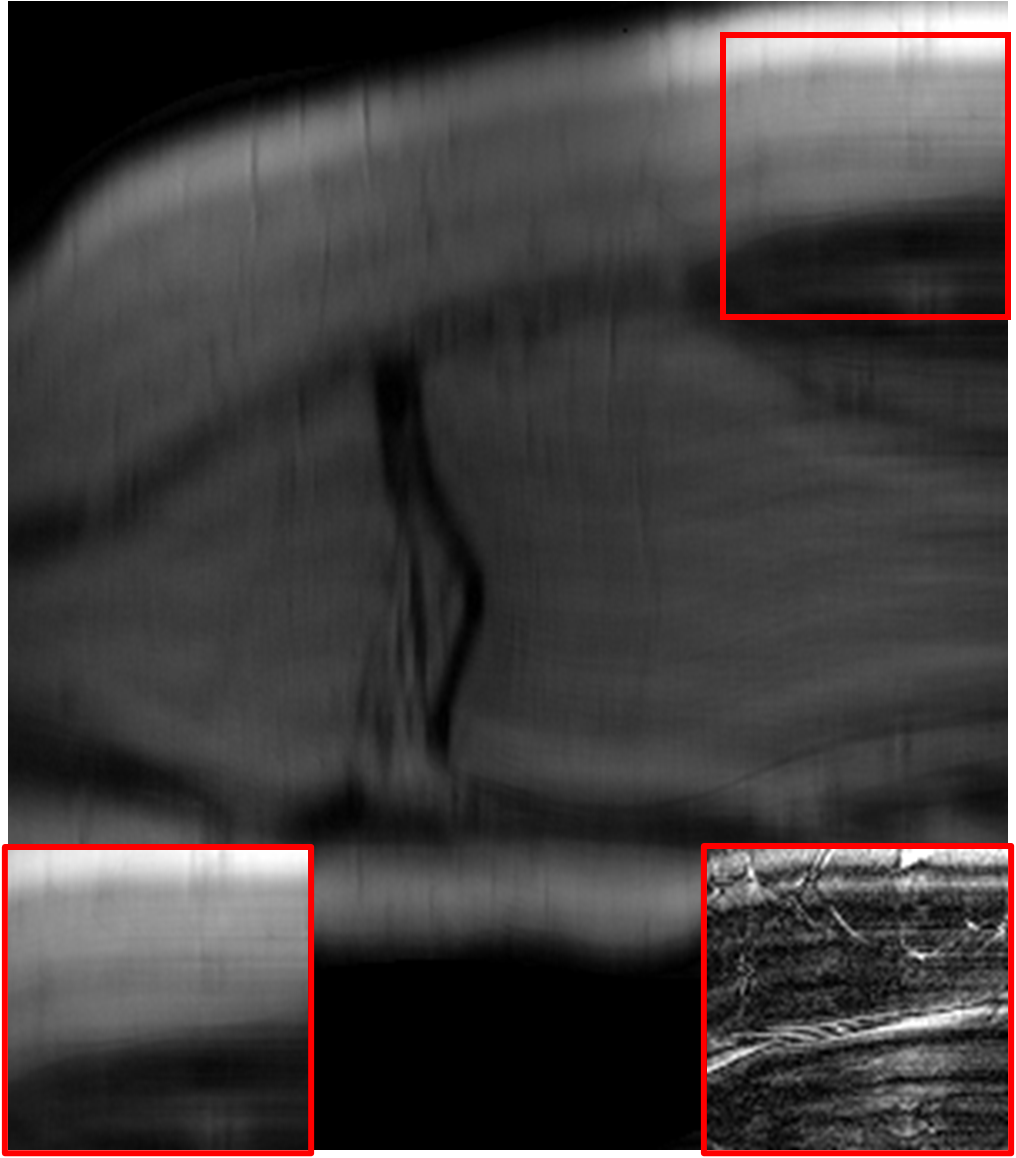}  \\
 (d) (34.20/0.957/0.621) \hspace{0.1in} & (e) (32.20/0.942/0.850) \hspace{0.1in} 
\end{tabular}
\caption{Comparison of reconstructions
for a knee image using the proposed method
versus strict supervised learning, blind dictionary learning, and zero-filled reconstruction
for the 5$\times$ undersampling mask depicted in \freff{fig:usml_msk}{a}. Metrics listed below each reconstruction correspond to PSNR/SSIM/HFEN respectively. \red{The inset panel on the bottom left in each image corresponds to regions of interest (indicated by the red bounding box in the image) in the image that benefits significantly from BLIPS reconstruction, while the inset on the bottom right depicts the corresponding error map.}}
\label{fig:recon:Fix_2}
\end{center}
\vspace{-0.1in}
\end{figure}

\renewcommand{\hh}{1.6in}
\begin{figure}[h!]
\begin{center}
\begin{tabular}{cc}
\textbf{Fully Sampled} &\\
\includegraphics[height=\hh]{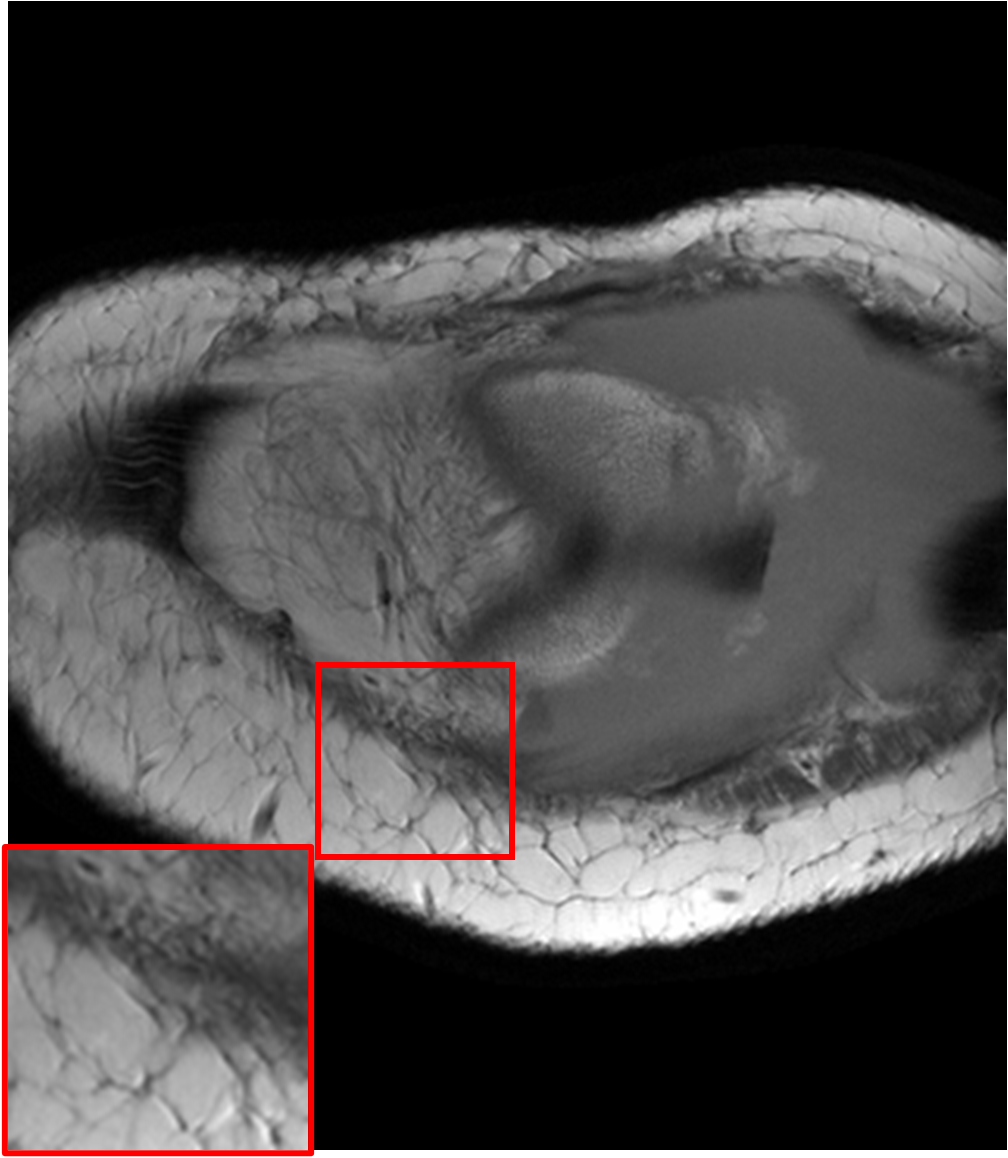} & \\ 
\hspace{0.1in}(a)(PSNR/SSIM/HFEN)\hspace{0.1in} &\\
\textbf{Blind+Supervised Recon.} & \textbf{Supervised Recon.} \\ 

\includegraphics[height=\hh]{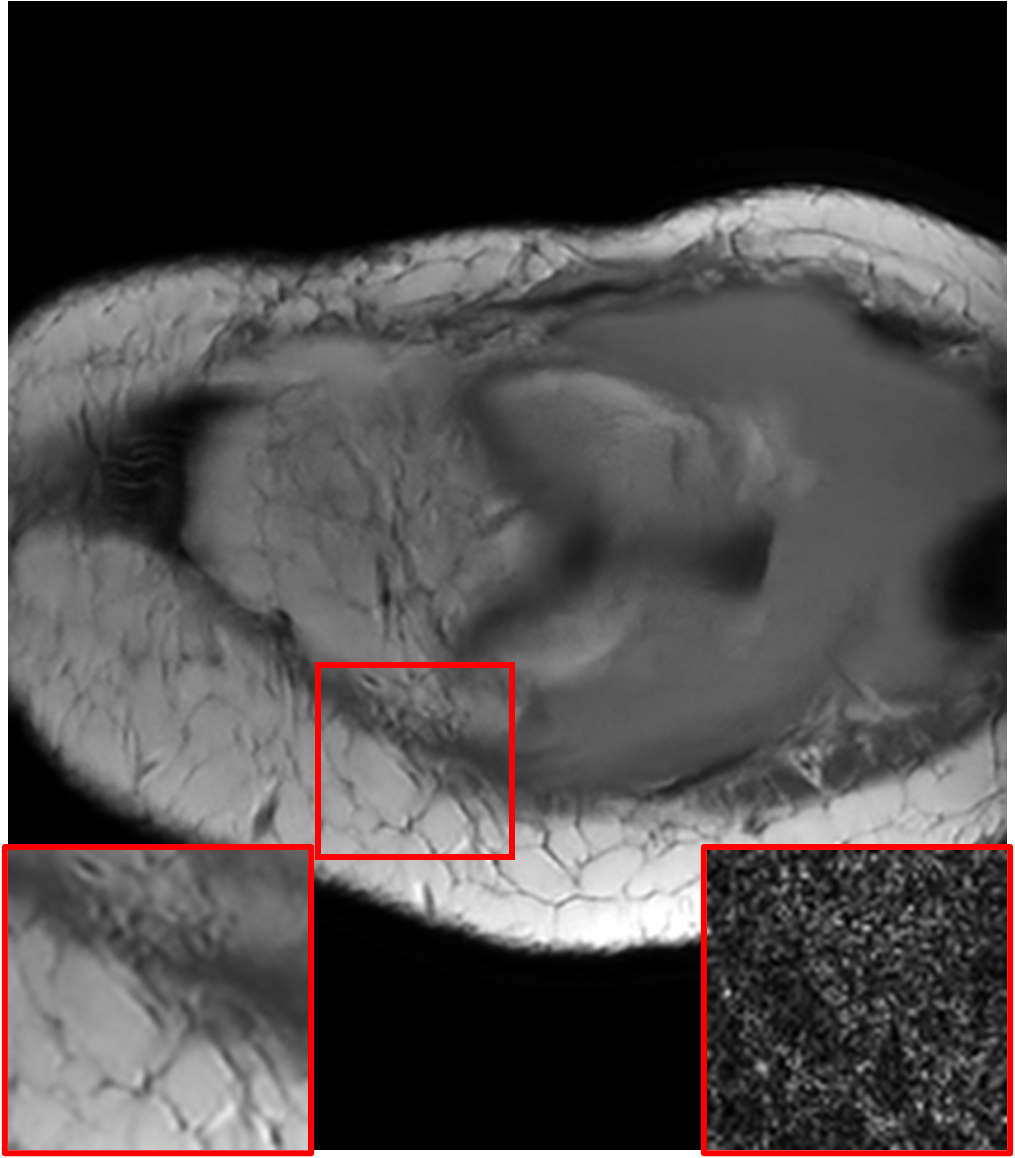}  & \includegraphics[height=\hh]{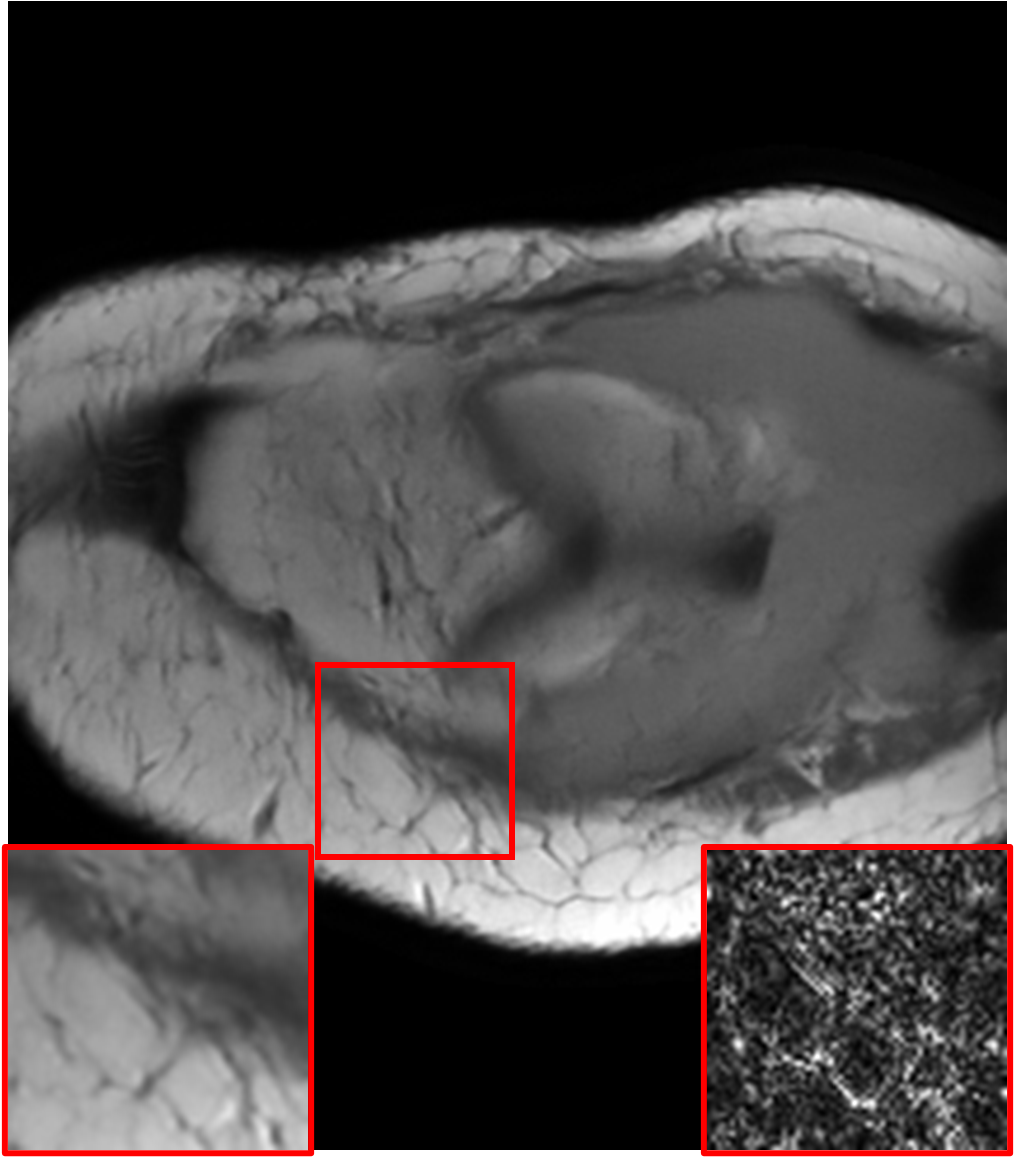}\\
 \hspace{0.1in}(b) (39.81/0.988/0.242) \hspace{0.1in} & \hspace{0.1in}(c) (37.84/0.984/0.329)\hspace{0.1in }\\
 \textbf{Blind Dict. Learning Recon.} & \textbf{Zero-Filled Recon.}\\
 \includegraphics[height=\hh]{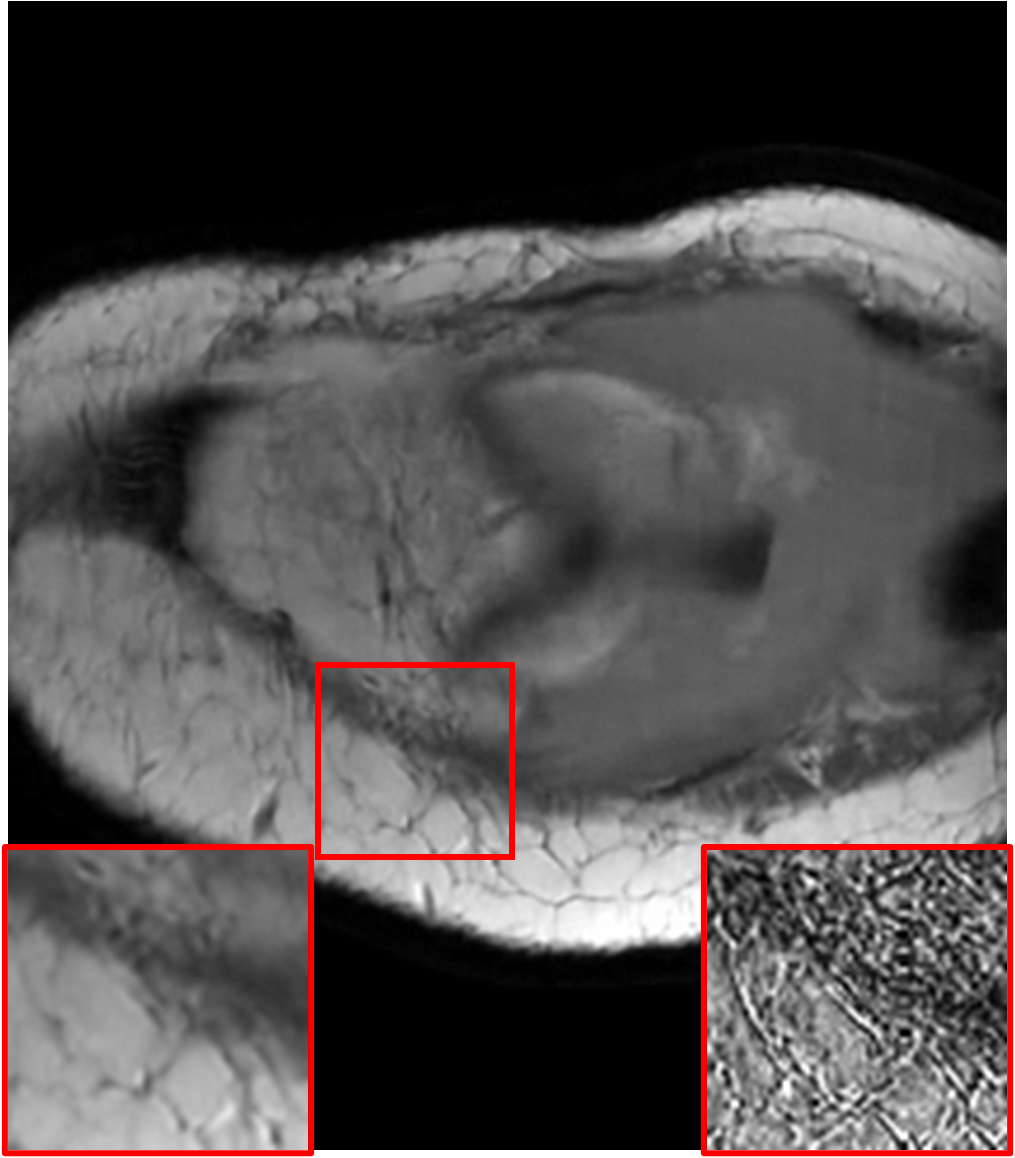}  & \includegraphics[height=\hh]{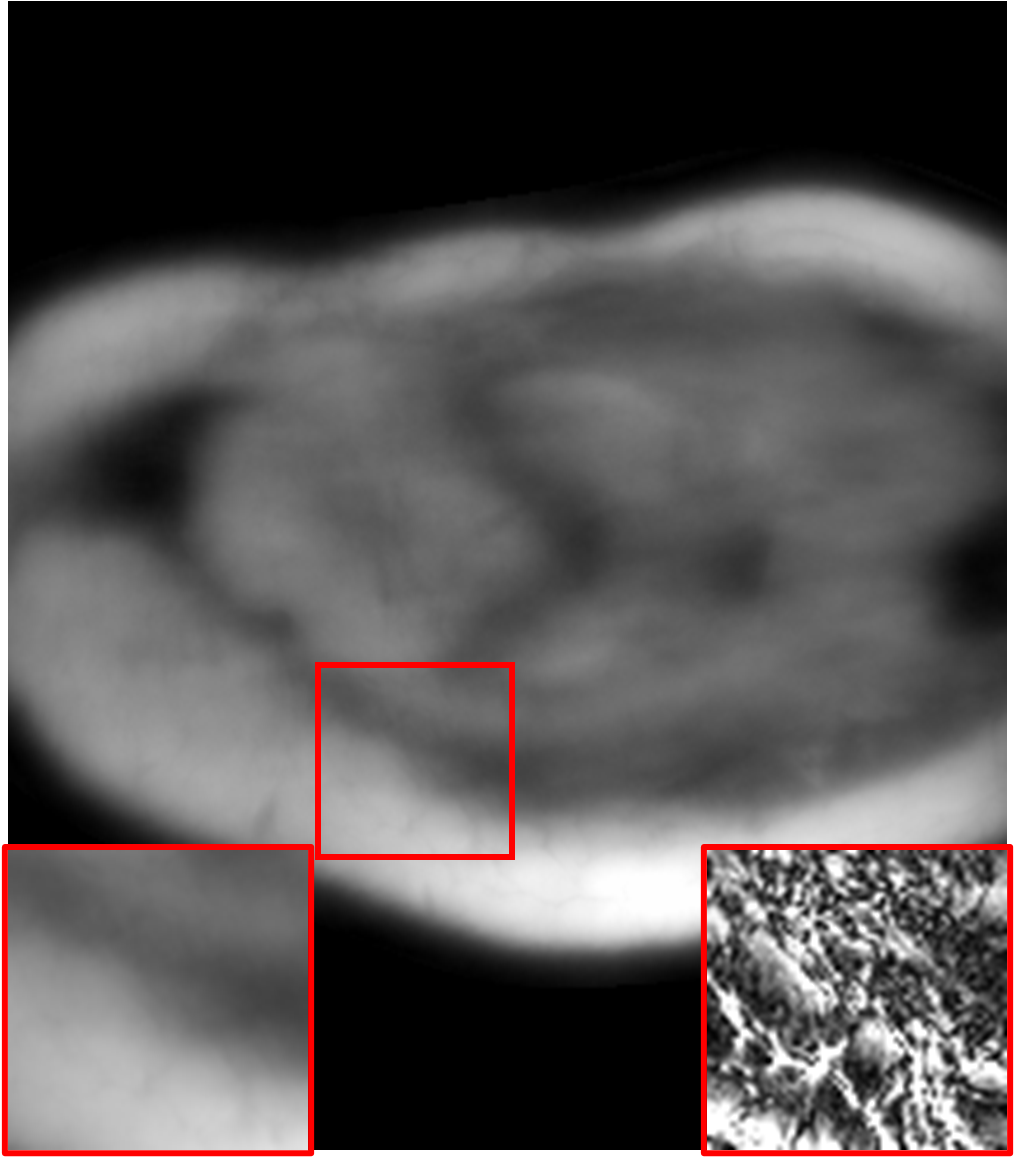}  \\
 (d) (32.86/0.981/0.984)\hspace{0.1in} & (e) (28.9/0.934/0.837) \hspace{0.1in} 
\end{tabular}
\caption{Comparison of reconstructions of a knee image using the proposed method
versus strict supervised learning, blind dictionary learning, and zero-filled reconstruction
for the 20$\times$ Poisson-disk undersampling mask depicted in \freff{fig:usml_msk}{b}. Metrics listed below each reconstruction correspond to PSNR/SSIM/HFEN respectively.
\red{The inset panel on the bottom left in each image corresponds to regions of interest (indicated by the red bounding box in the image) in the image that benefits significantly from BLIPS reconstruction, while the inset on the bottom right depicts the corresponding error map.}
}
\label{fig:recon:Psn_2}
\end{center}
\vspace{-0.1in}
\end{figure}

\renewcommand{\hh}{1.6in}
\begin{figure}[h!]
\begin{center}
\begin{tabular}{cc}
\textbf{Fully Sampled} & \\ 
\includegraphics[height=\hh]{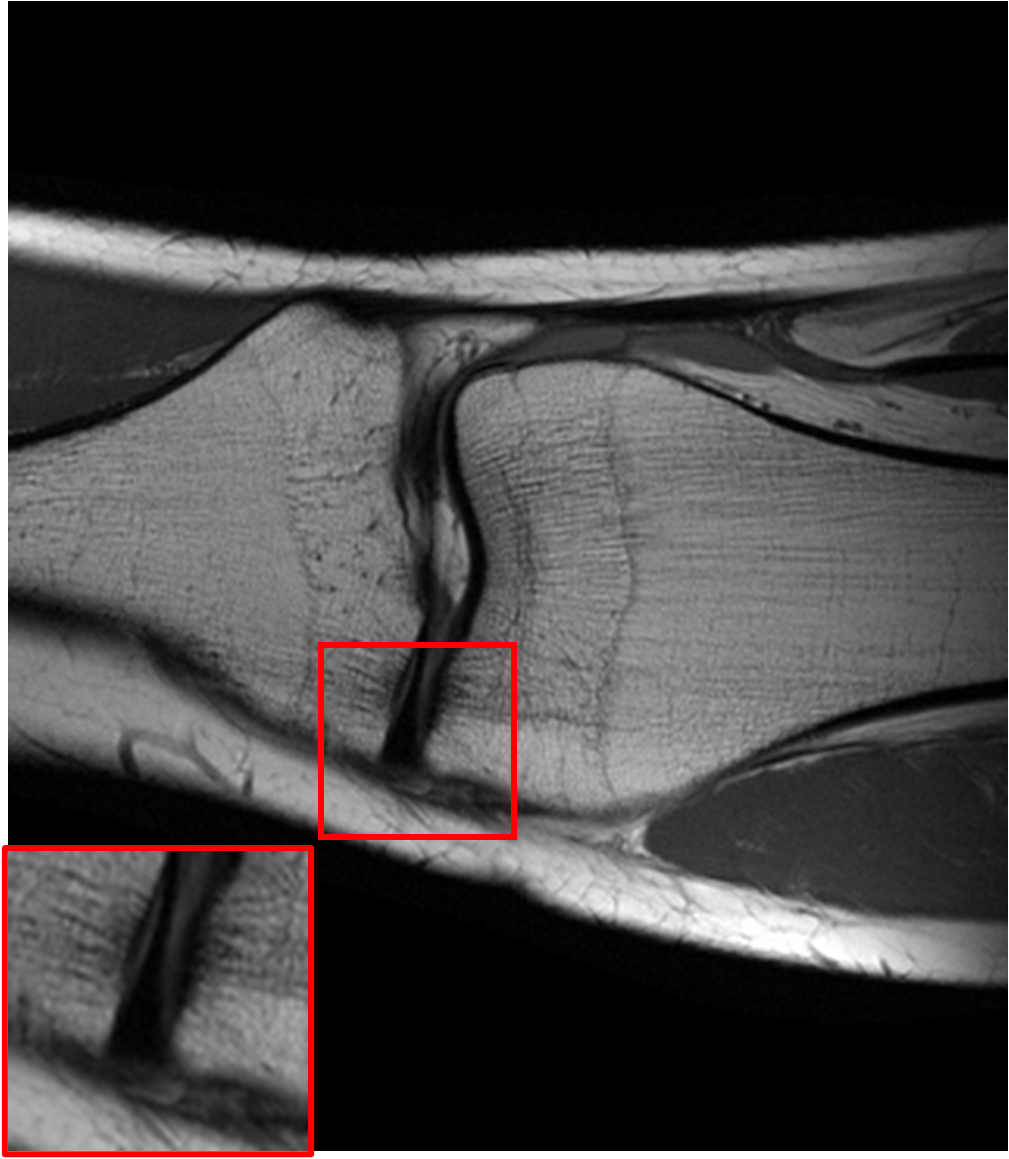} & \\
\hspace{0.1in}(a)(PSNR/SSIM/HFEN)\hspace{0.1in} &\\
\textbf{Blind+Supervised Recon.} & \textbf{Supervised Recon.} \\ 
 \includegraphics[height=\hh]{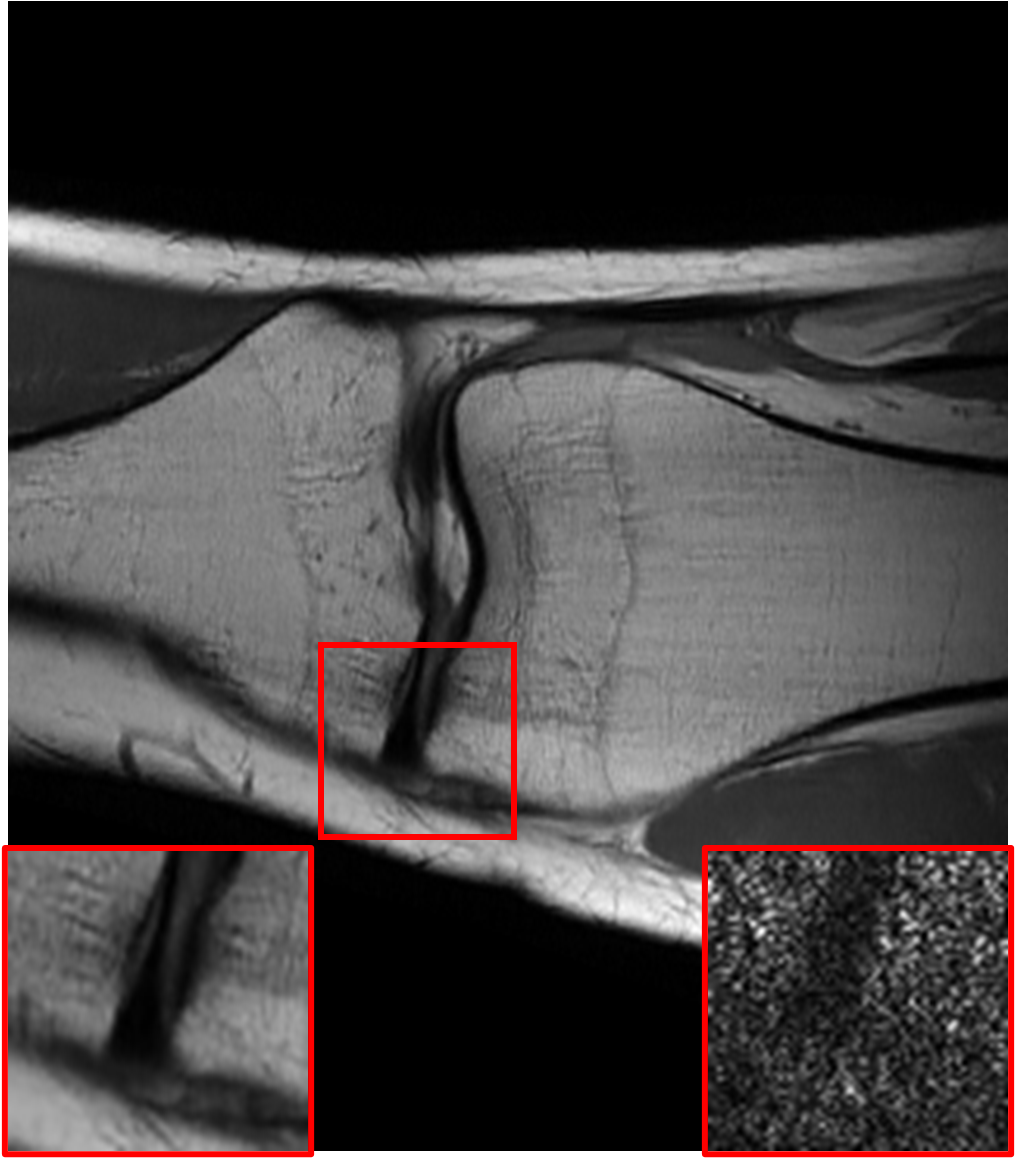}  & \includegraphics[height=\hh]{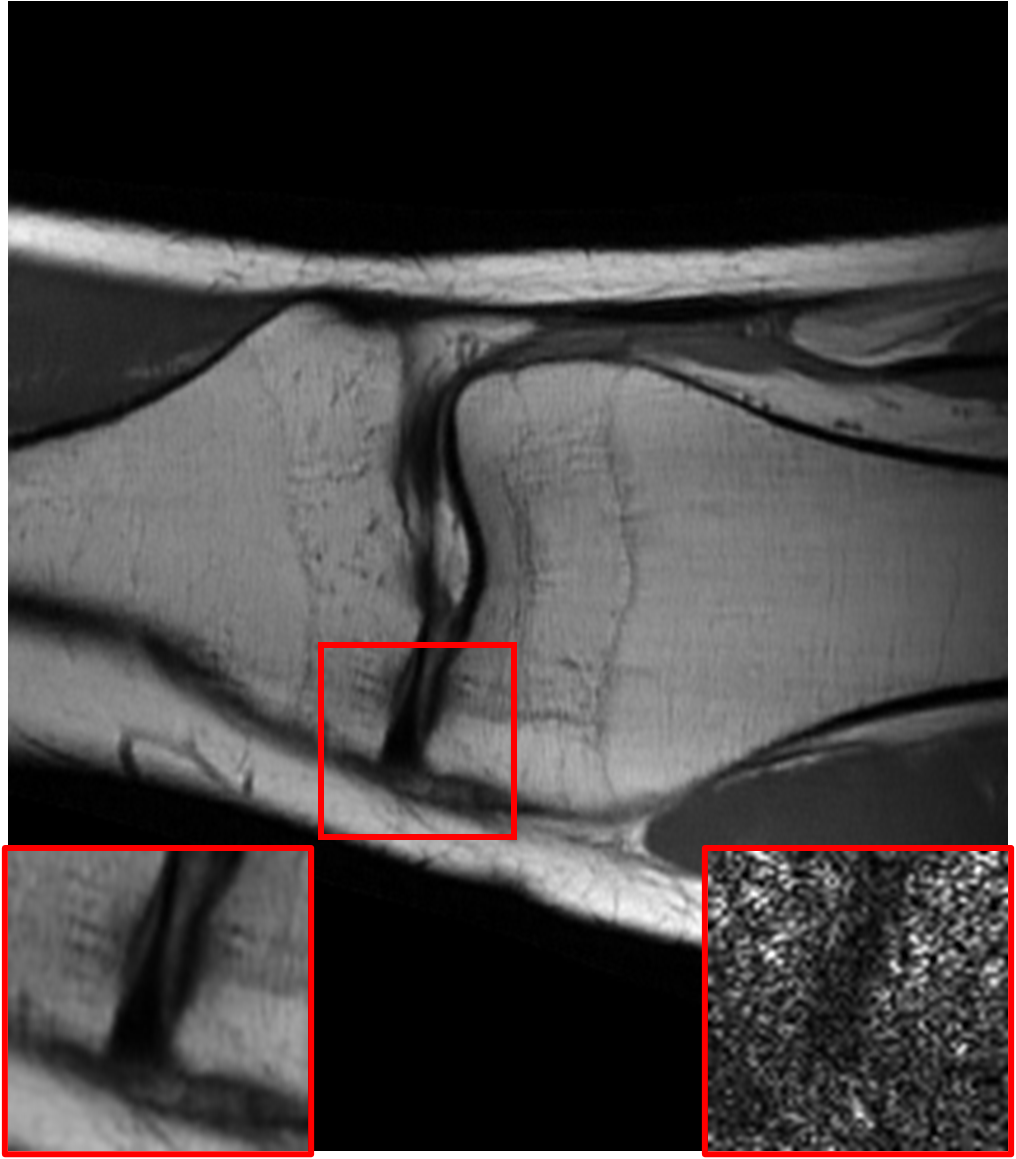}\\
 \hspace{0.1in}(b) ( 35.60/0.970/0.164) \hspace{0.1in} & \hspace{0.1in}(c) (34.66/ 0.964/0.209) \hspace{0.1in } \\
\textbf{Blind Dict. Learning Recon.} & \textbf{Zero-Filled Recon.}\\
 \includegraphics[height=\hh]{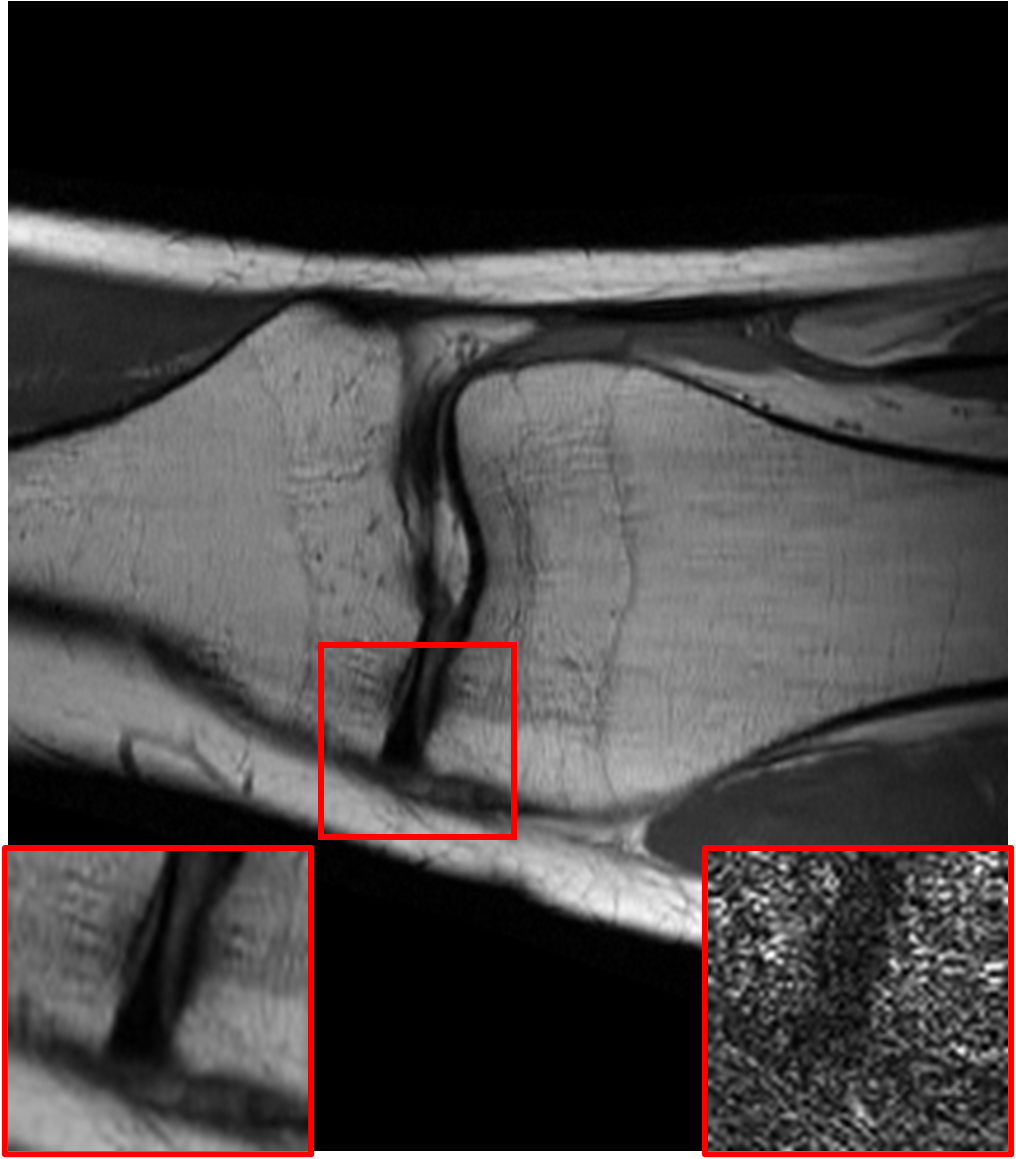}  & \includegraphics[height=\hh]{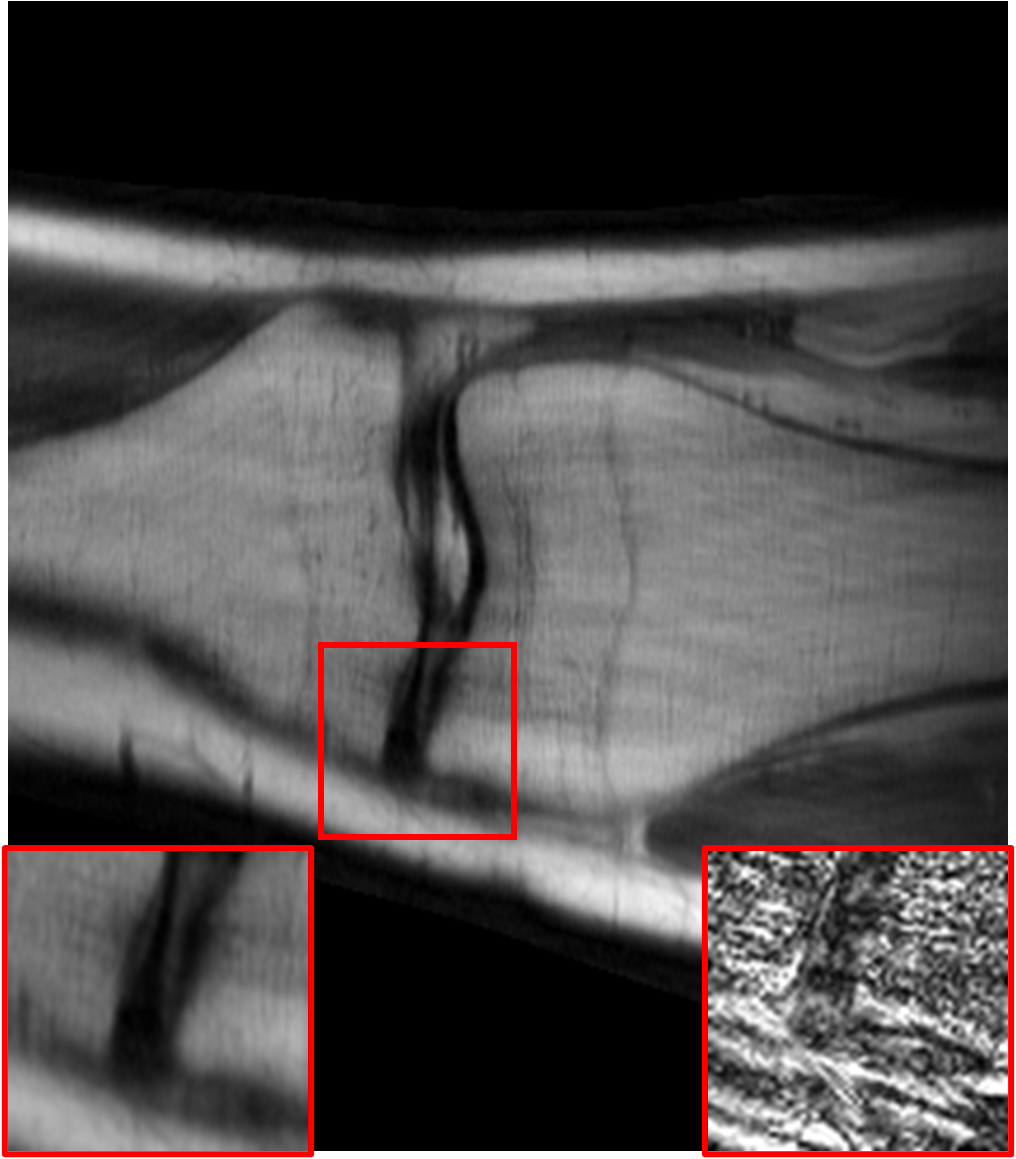}  \\
 (d) (33.75/0.962/0.219)\hspace{0.1in} & (e) (27.66/0.881/0.628) \hspace{0.1in}
\end{tabular}
\caption{Comparison of reconstructions of a knee image using the proposed method versus strict supervised learning, blind dictionary learning, and zero-filled reconstruction for the random 1D undersampling masks ($\approx  4.5\times$). Metrics listed below each reconstruction correspond to PSNR/SSIM/HFEN respectively.
\red{The inset panel on the bottom left in each image corresponds to regions of interest (indicated by the red bounding box in the image) in the image that benefits significantly from BLIPS reconstruction, while the inset on the bottom right depicts the corresponding error map.}}
\label{fig:recon:Rnd_2}
\end{center}
\vspace{-0.1in}
\end{figure}

\renewcommand{\hh}{1.6in}
\begin{figure}[h!]
\begin{center}
\begin{tabular}{cc}
\textbf{Fully Sampled} & \\ 
\includegraphics[height=\hh]{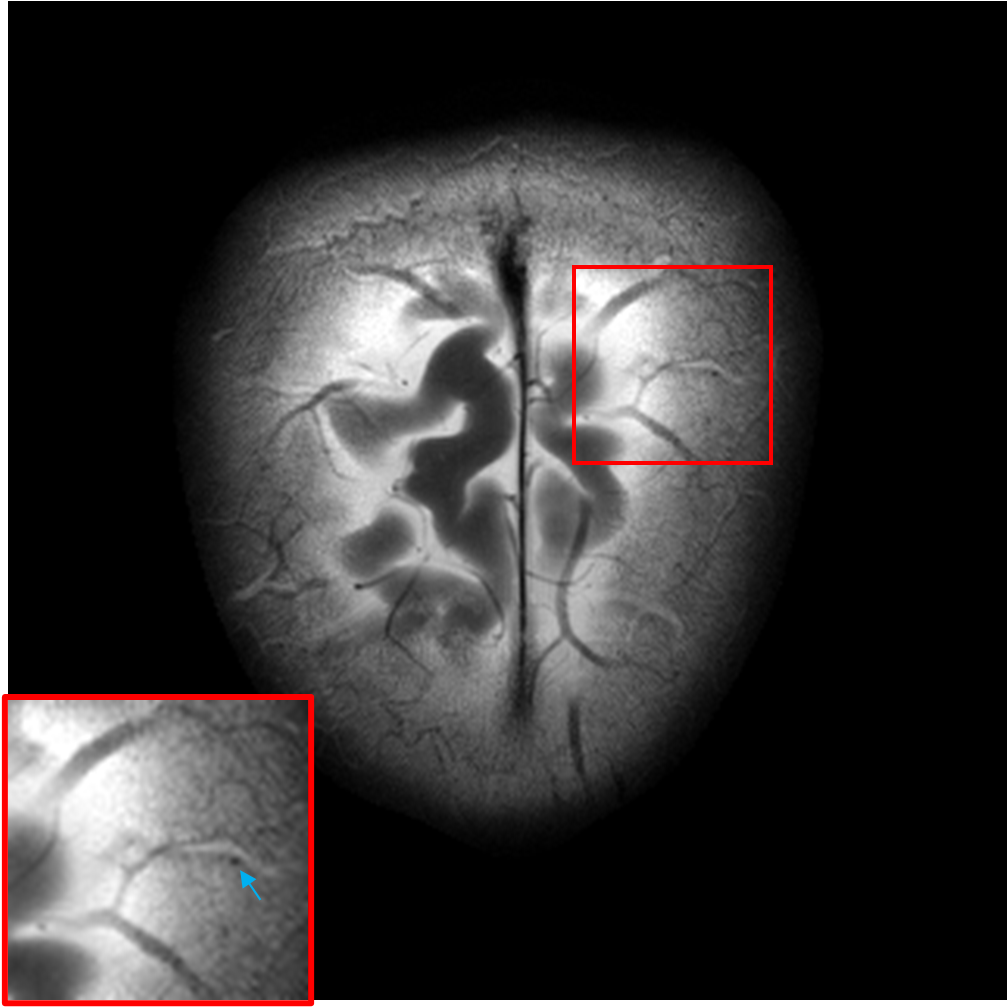} & \\
\hspace{0.1in}(a)(PSNR/SSIM/HFEN)\hspace{0.1in} &\\
\textbf{S+B+S Recon.} & \textbf{S+S Recon.} \\ 
 \includegraphics[height=\hh]{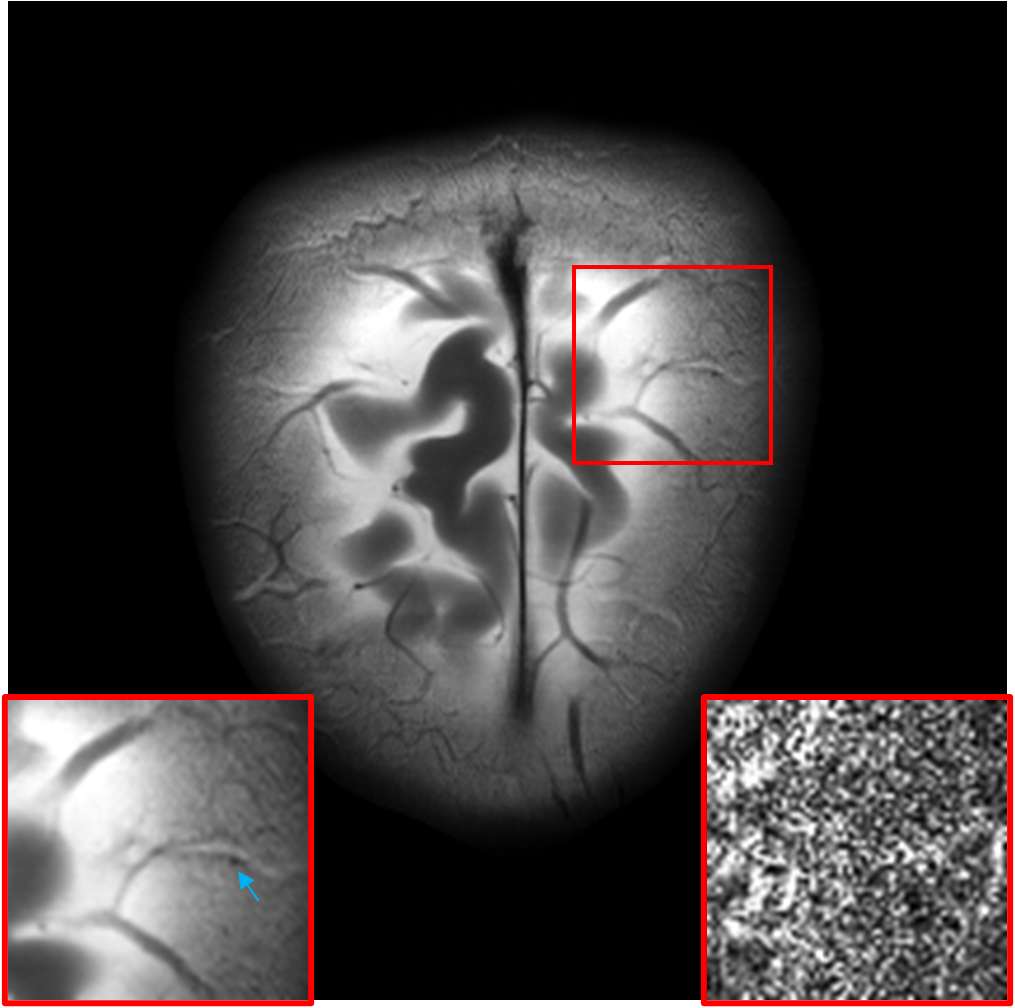}  & \includegraphics[height=\hh]{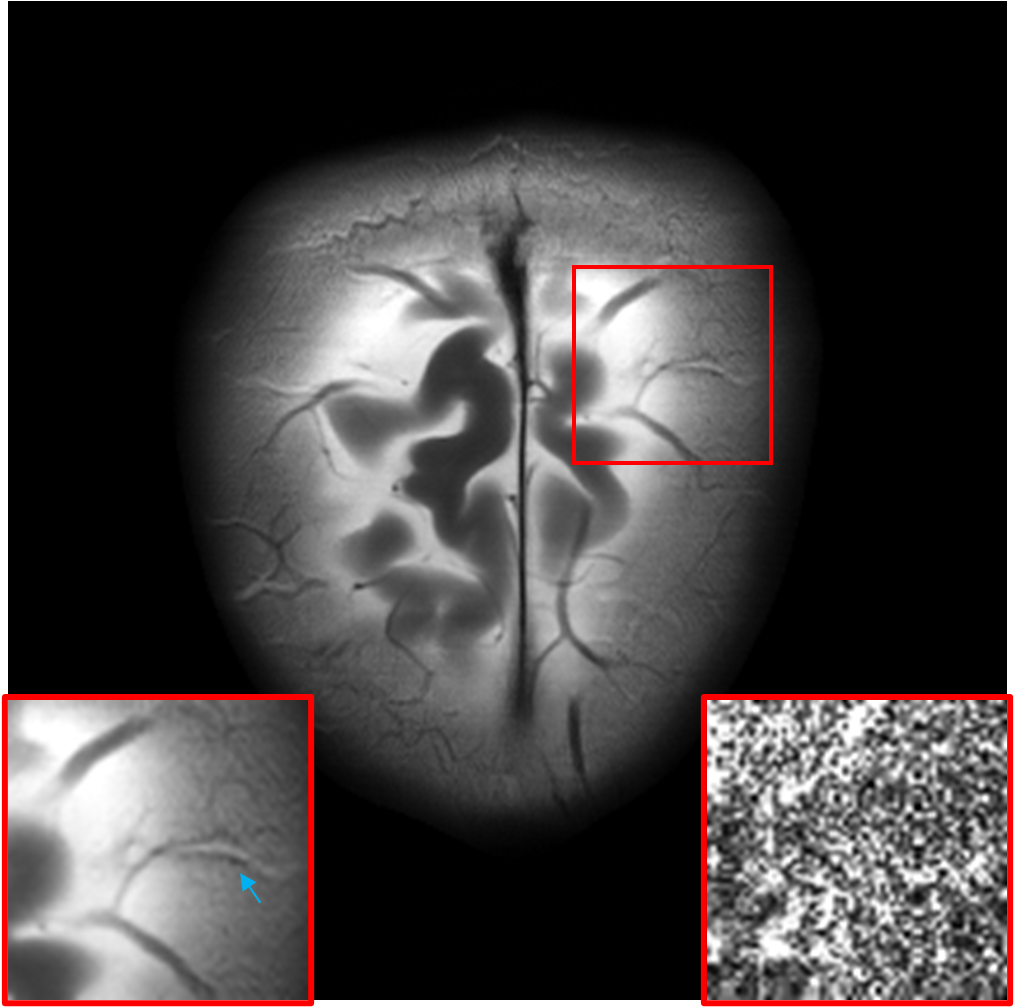}\\
 \hspace{0.1in}(b) (40.00/0.987/0.269) \hspace{0.1in} & \hspace{0.1in}(c) (38.73/0.983/0.320) \hspace{0.1in }\\
 %\hspace{0.1in}(b) (30.07/0.884/0.487) \hspace{0.1in} & \hspace{0.1in}(c) (29.71/0.878/0.516) \hspace{0.1in }\\
\textbf{S+B Recon.} & \textbf{Zero-Filled Recon.}\\
 \includegraphics[height=\hh]{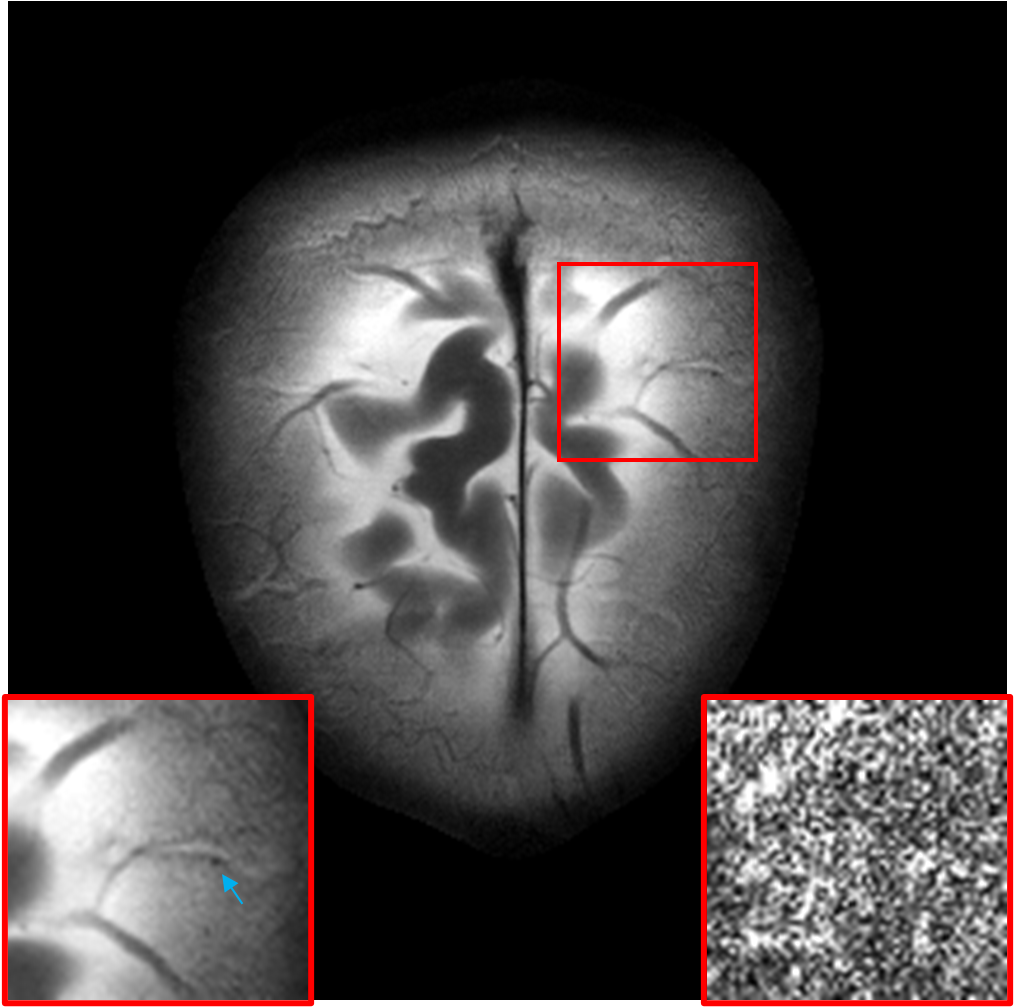}  & \includegraphics[height=\hh]{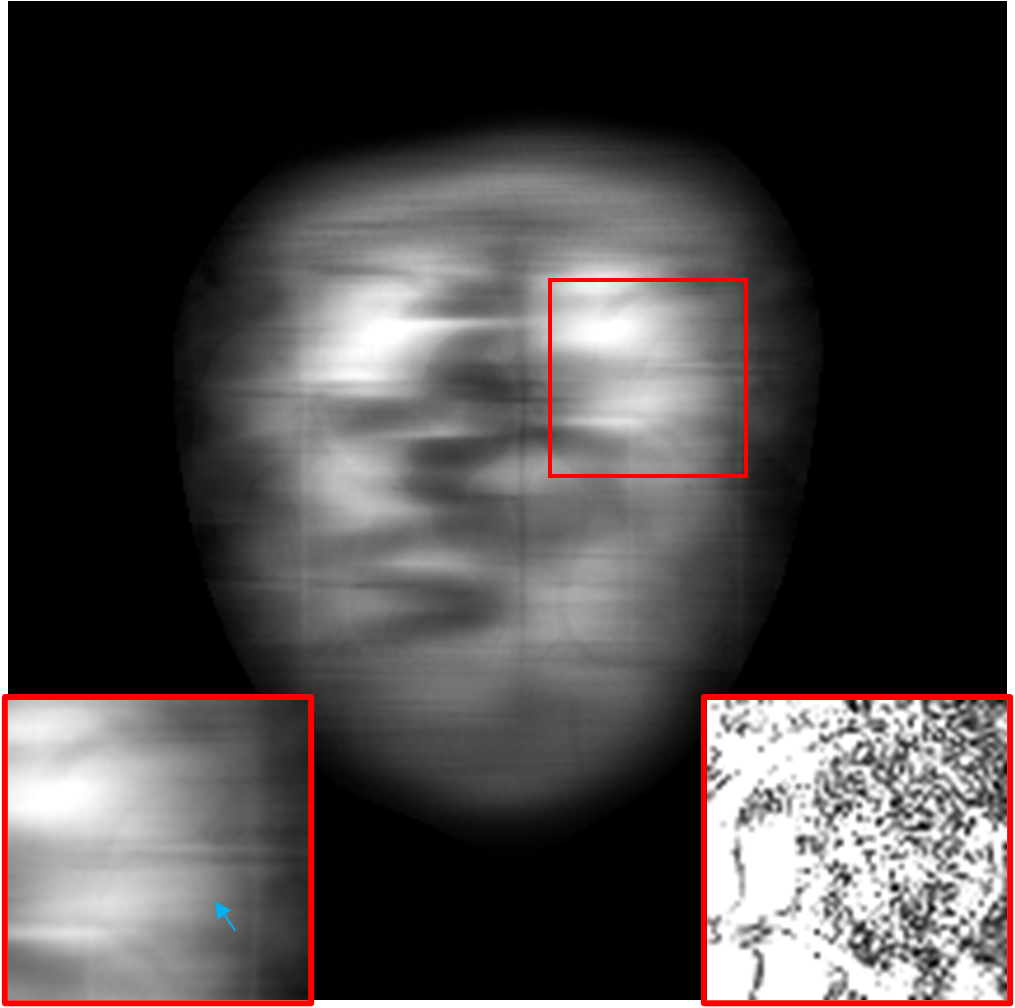}  \\
 \hspace{0.1in}(d) (38.89/0.981/0.314)\hspace{0.1in} & \hspace{0.1in}(e) (30.13/0.9451/0.8615) \hspace{0.1in}
\end{tabular}
\caption{{Comparison of reconstructions for two T2w brain images using the S+B+S learning reconstruction method proposed in (P3) versus cascaded S+S strict supervised learning-based reconstruction, S+B reconstruction, and zero-filled reconstruction for an 8$\times$ equidistant undersampling mask. The S+B reconstruction depicts the output of one iteration of blind reconstruction initialized with a supervised reconstruction. Metrics listed below each reconstruction correspond to PSNR/SSIM/HFEN respectively.}\red{The inset panel on the bottom left in each image corresponds to regions of interest (indicated by the red bounding box in the image) in the image that benefits significantly from BLIPS reconstruction, while the inset on the bottom right depicts the corresponding error map. The blue arrows indicate the position of image detail that is present in the the BLIPS reconstruction, but not strict supervised learning-based reconstruction.}
}
\label{fig:recon:equi_8x_brain}
\end{center}
\vspace{-0.1in}
\end{figure}
\newcommand{\hhh}{2.2in}

\begin{figure*}[b]
\begin{center}
\begin{tabular}{ccc}
\textbf{Fully Sampled} & \textbf{Blind+Supervised Recon.} & \textbf{Supervised Recon.} \\
\includegraphics[height=\hhh]{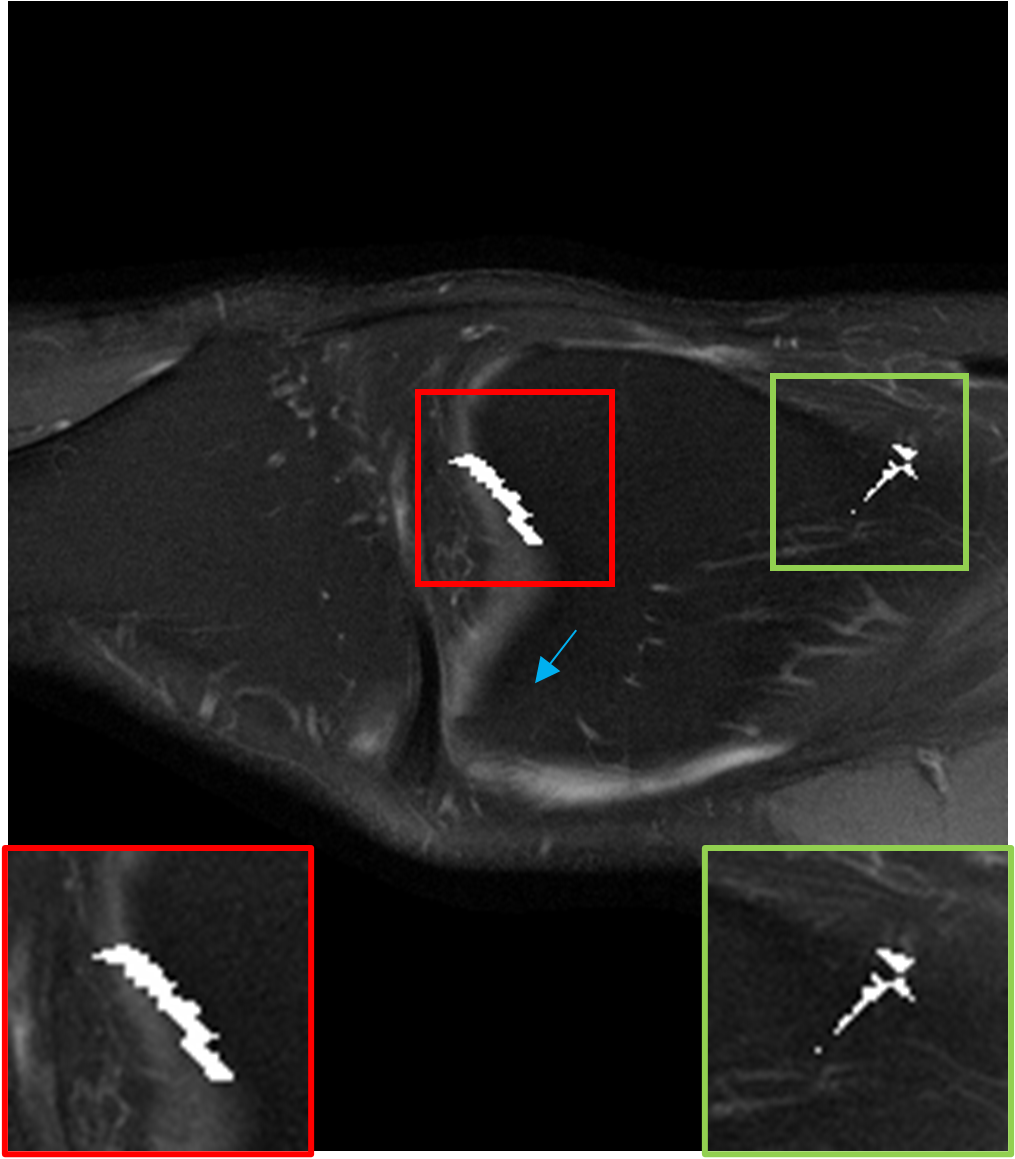} & \includegraphics[height=\hhh]{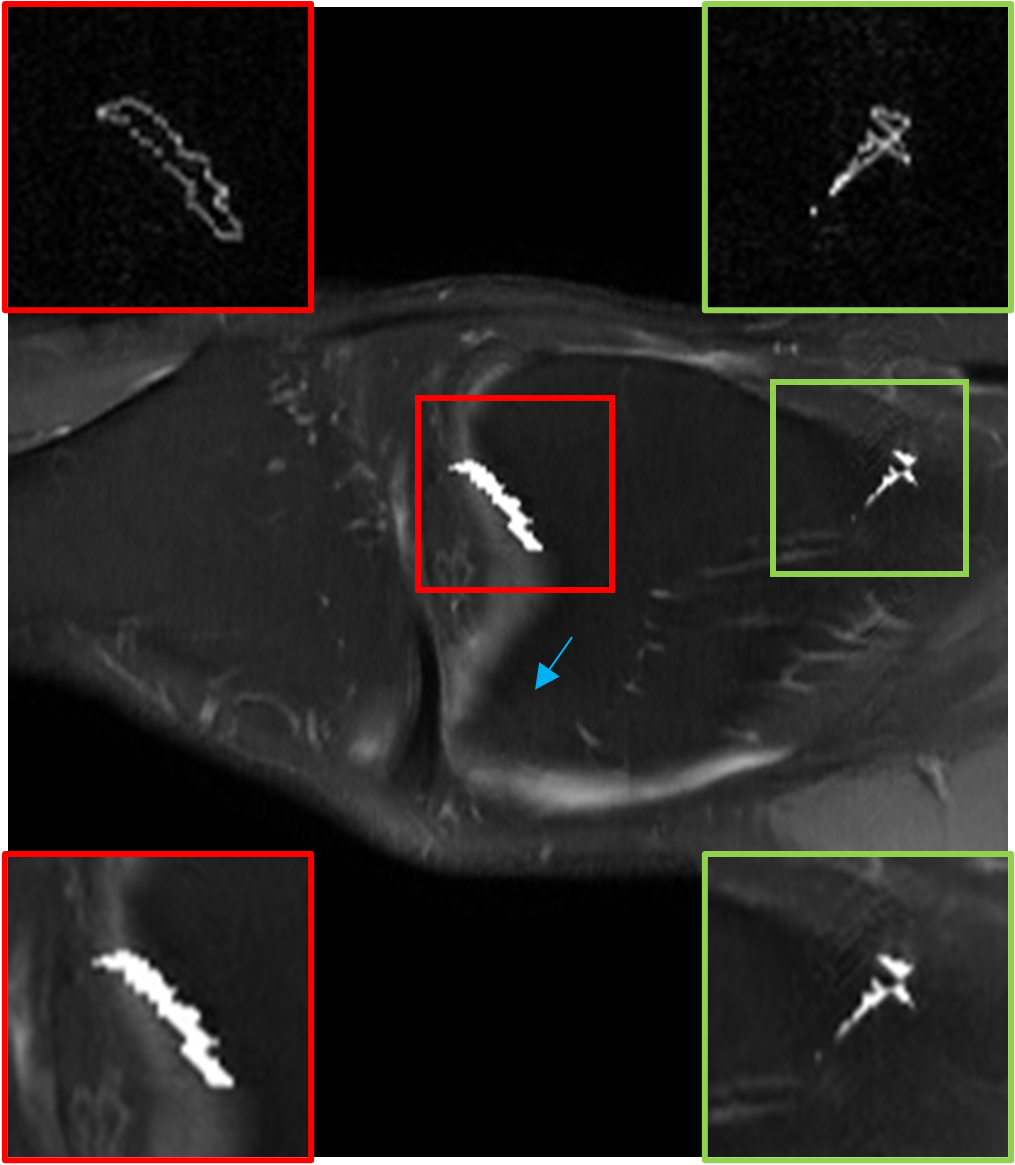}  & \includegraphics[height=\hhh]{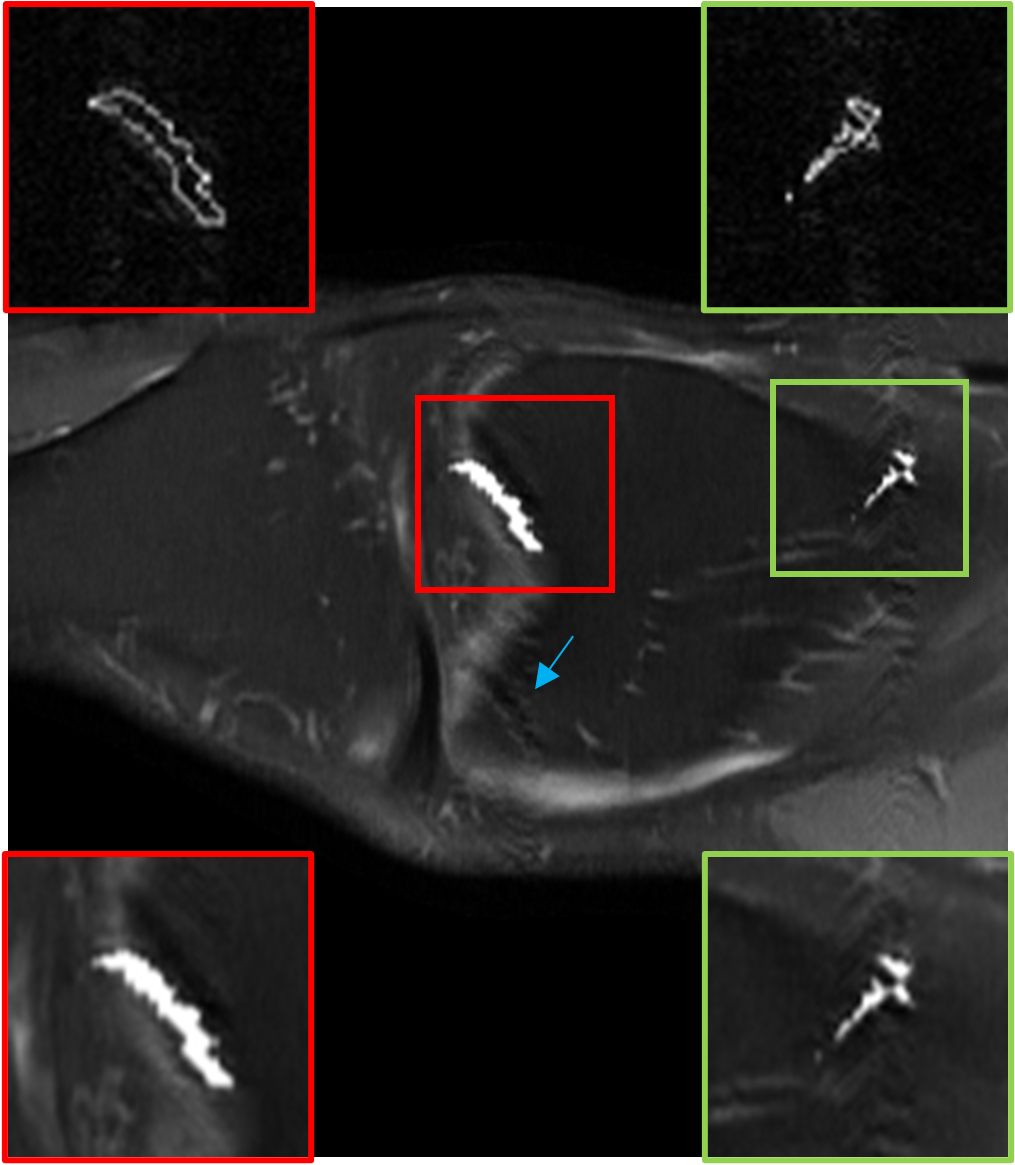}\\ 
\hspace{0.1in}(a)(PSNR/SSIM/HFEN)\hspace{0.1in} & \hspace{0.1in}(b) (35.00/0.960/0.2263) \hspace{0.1in} & \hspace{0.1in}(c) (33.83/0.955/0.3020)\hspace{0.1in }\\
 & \textbf{Blind Dict. Learning Recon.} & \textbf{Zero-Filled Recon.}\\
 & \includegraphics[height=\hhh]{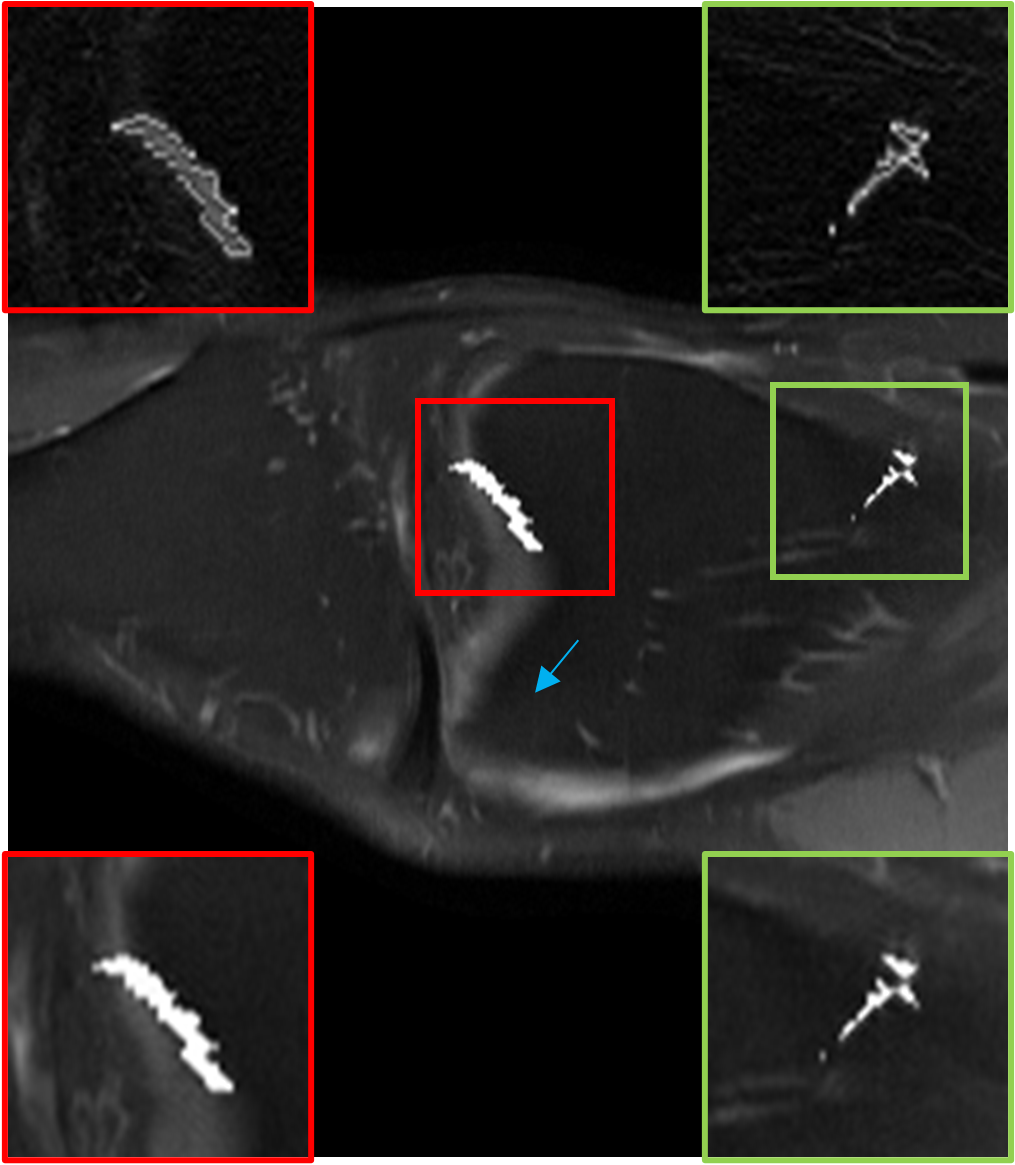}  & \includegraphics[height=\hhh]{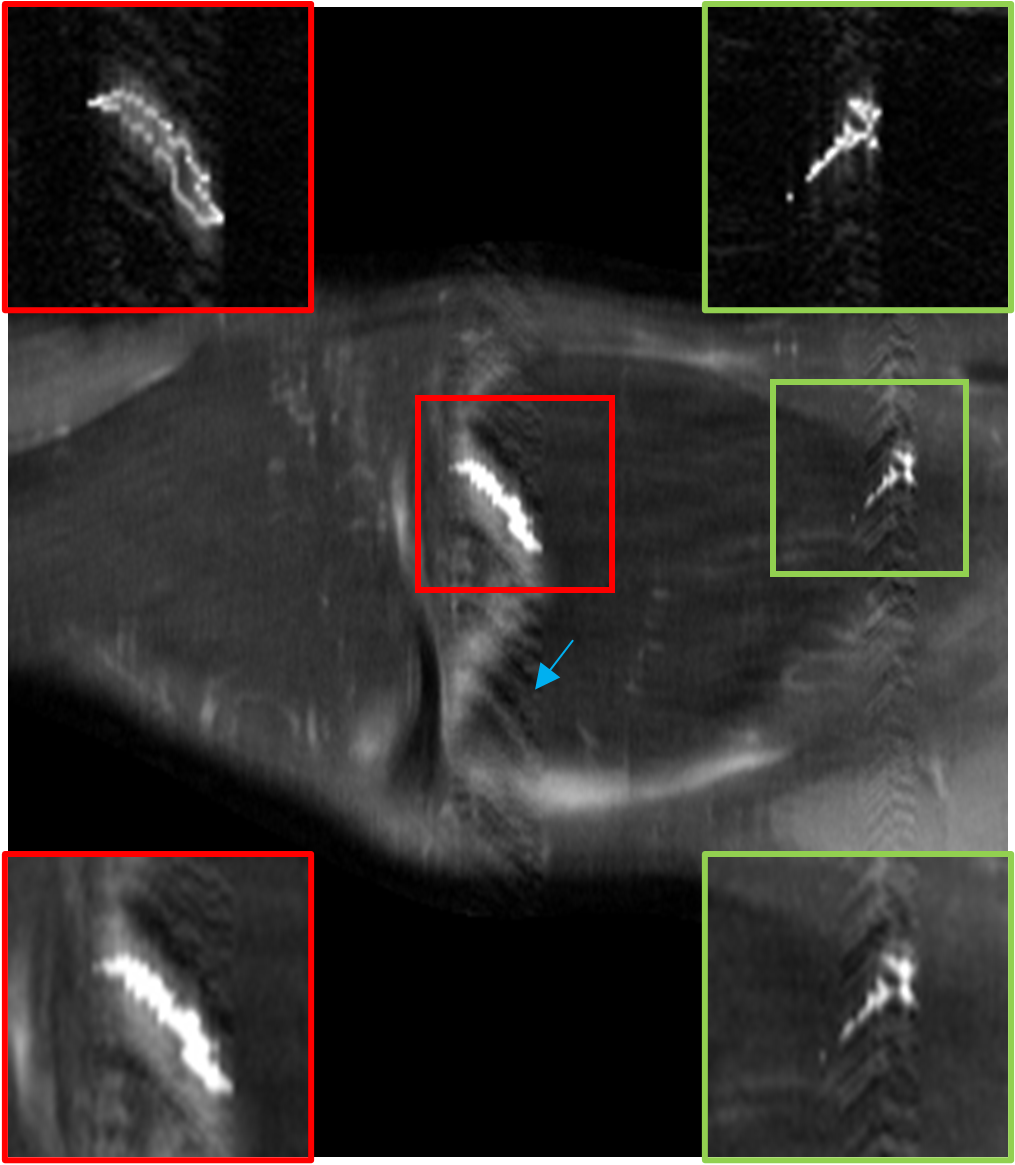}  \\
& (d) (29.86/0.948/0.2985)\hspace{0.1in} & (e) (30.07/0.915/0.6356) \hspace{0.1in} 
\end{tabular}

\caption{Comparison of reconstructions of a knee image using the proposed method
versus strict supervised learning
for an image slice with artificially planted features. The undersampling mask was chosen to be random $\approx 4.5\times$. Metrics listed below each reconstruction correspond to PSNR/SSIM/HFEN respectively.
\red{The inset panels on the bottom in each image correspond to regions of interest (indicated by the red/green bounding boxes in the image) in the image that benefit significantly from BLIPS reconstruction, while the insets on the top depicts the corresponding error map. The blue arrows indicate the position of an aliasing artifact that is present in the zero-filled reconstruction and strict supervised learning, but not in the BLIPS reconstruction.}}
\label{fig:recon:patho}
\end{center}
\vspace{-0.1in}
\end{figure*}

%% file: s,conclusion.tex
% s,conclusion

{
This paper investigated a combination of shallow dictionary learning and deep supervised learning
for MR image reconstruction
that leverages the complementary nature of the two methods
to bolster the quality of the reconstructed image.
We verify this benefit by comparisons
using a variety of metrics (including SSIM, PSNR, and HFEN)
against strictly supervised learning-based reconstruction,
reconstruction as initialization.
We also investigate alternative approaches
for combining the two forms of reconstruction.
Our observations suggest that the primary benefits
of including blind learning in the reconstruction pipeline
are the preservation of `finer' details in the output image
and robustness to the availability of training data.
}

In the future,
we aim to apply our methods to non-Cartesian undersampling patterns such as radial and spiral patterns,
and to other modalities.
The generalizability of the method, especially with heterogeneous datasets, will be further explored.
{We observed some variation in the performance of our method
to the imposed sparsity level in \eqref{eq:dl}.
More careful tuning of hyperparameters will be necessary
to optimize
the overall performance of such methods.
Curiously, we also observed that using additional iterations of blind learning reconstruction
in \eqref{SBS}
adversely impacted the performance of our methods.
The cause for this behavior is unknown (beyond oversmoothing),
and needs further investigation.
}
We also plan to investigate the benefits of multiple iterations
of combined blind and supervised learning based reconstruction,
extending the S+B+S approach considered here.
\resp{3.5}\red{Aside from the benefits 
of traditional `handcrafted' priors
in combination with supervised deep learning,
from the perspective of learning 
only from measurements of the image
being reconstructed,
and then filling in the gaps with supervised data-driven learning,
it would be interesting to study 
the combination of deep blind approaches%
\cite{Oh2020UnpairedDeepLearning,yaman2020self,tamir2019unsupervised,ke2020unsupervised}
with deep supervised learning.}

%% file: sm.tex
\subsection{Comparison with Non-Adaptive Dictionary-based Initialization for Supervised Learning}
\label{sec:fixdct}
\resp{4.4}In this experiment,
we fixed the dictionary \D in \eqref{eq:dl}
as an overcomplete inverse DCT matrix
and did not update it.
%and skipped updating \D.
We then used the resulting reconstructed image
to initialize the supervised reconstruction algorithm.
Table VI % i think this is the wrong table #. it is very risky to use "VI" instead of \ref{}
% now you must manually check every table reference to make sure you have it correct.
compares the results of proposed blind+supervised learning
versus non-adaptive dictionary-based initialization for supervised learning.
The sampling pattern remains the same as in the previous case.
\red{4198} slices were used for training, with 10\% left for validation.
The test set consisted of \red{500} slices.
%\end{multicols}
\begin{table}[h!]
    \centering
    \begin{tabular}{|c||c|c|}
    \hline
        Recon. Method & Fixed Dictionary+Supervised & Blind+Supervised \\\hline\hline
         SSIM & 0.945  &  {0.946} \\
         PSNR (dB) &  35.37 & {35.53}\\
         HFEN &  0.452 & {0.443} \\
         \hline
    \end{tabular}
    \begin{tabular}{ccc}
        \multicolumn{3}{l}{} \\
    \multicolumn{3}{l}{\textbf{Table VII}: Comparison of performance of
    non-adaptive dictionary-based  } \\
     \multicolumn{3}{l}{initialization for supervised learning-based reconstruction
     versus our proposed } \\
     \multicolumn{3}{l}{combined blind and supervised learning-based reconstruction,
     for the  } \\
     \multicolumn{3}{l}{undersampling mask shown in \freff{fig:usml_msk}{b}. The data set involved is the 4198/500}\\
     \multicolumn{3}{l}{slices from the fastMRI knee dataset.}
    \end{tabular}
    \label{tab:CS}
\end{table}

\red{We surmise that the reason for relatively small improvements with blind learning 
over a fixed dictionary initialization in our proposed pipeline is due to the lack of 
proper parameter tuning during dictionary learning.
One way to remedy this would be to vary the sparsity penalty weight, $\lambda$,
across outer-iterations of dictionary learning-based reconstruction
as is done in \cite{Ravishankar2017EfficientProblems}.
Furthermore, the initialization for blind dictionary learning 
was a zero-filled reconstruction, which can be detrimental 
to learning a `good' dictionary. We expect that addressing 
these issues could further bolster the performance of BLIPS reconstruction.}

\subsection{Contribution of Residual Supervised Learning}

\begin{figure}[h!]
    \centering
    \includegraphics[width=3.0in]{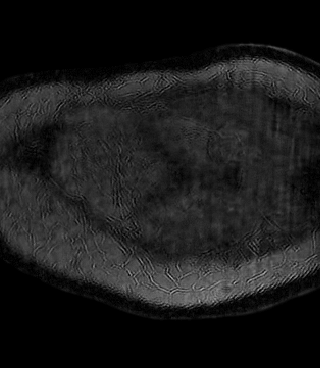}
    \caption{Residual contribution of the supervised learning module for the image in \fref{fig:recon:Psn_2}, obtained by removing the blind dictionary learning output from the BLIPS reconstructed image.}
    \label{fig:residue}
\end{figure}

\red{To gain more insight
\resp{2.1,2.2}
into the 
mechanism of the proposed BLIPS reconstruction, 
we examined the residual component 
added to the blind dictionary learning-based reconstruction
by the supervised learning-based reconstruction component. 
Essentially,
we removed the blind learning output 
from the BLIPS reconstruction
to study the contribution of the supervised module.
\fref{fig:residue} shows the contribution of the
supervised learning component for \fref{fig:recon:Psn_2}.}

\red{We observe that the supervised learning module
mainly contributes to removing left-over aliasing 
artifacts from the blind learning-based reconstruction, 
and also focuses on sharpening the details in the blind reconstruction. 
This observation reinforces the concept of complementarity of blind and supervised learning-based reconstruction.}

\subsection{Reconstruction Times}

\red{Table VIII
\resp{3.3,4.10}
lists the reconstruction times for the various methods proposed and compared to in this work.
Strict supervised learning is the fastest,
while the BLIPS approach in \fref{fig:pipeline} (P1) is the slowest,
because it requires several iterations of the SOUP-DIL algorithm \cite{Ravishankar2017EfficientProblems},
currently implemented in Matlab.
This drawback may be remedied by providing a better initialization for dictionary learning
and using GPUs for acceleration.
}
\begin{table}[h!]
    \centering
    \begin{tabular}{|c|c|}
    \hline
         Recon Method & Recon Time (s) \\
         \hline\hline
         S & 1.2 \\
         B & 170\\
         CS+S &  {80.2}\\
         B+S & 171.2\\
         S+B+S & 8.7\\
         \hline
    \end{tabular}
    \begin{tabular}{ccc}
        \multicolumn{3}{l}{} \\
    \multicolumn{3}{l}{\textbf{\red{Table VIII}}: Comparison of reconstruction times of various  } \\
     \multicolumn{3}{l}{methods explored in our work } \\
    \end{tabular}
    \label{tab:recon_times}
\end{table}

%\subsection